\documentclass[11pt]{article}
\usepackage{amsmath}\usepackage{amsfonts}
\usepackage{CJK}
\usepackage{subfigure}
\usepackage{graphicx}
\usepackage{dsfont}
\usepackage{color}
\usepackage{amscd}
\newtheorem{thm}{Theorem}

 \newtheorem{prop}{Proposition}
\newtheorem{rem}{Remark}
\begin{document}

\title{High Order Solutions  and Generalized Darboux Transformations of Derivative Schr\"odinger Equation}
\author{Boling Guo$^\dag$, Liming Ling$^{*,\dag}$ and Q. P. Liu$^\ddag$\\
$^\dag$Institute of Applied Physics and Computational Mathematics\\
Beijing 100088, P R China\\[10pt]
$^*$The Graduate School of China Academy of Engineering Physics,\\
Beijing 100088, P R China\\[10pt]
and\\
$^\ddag$Department of Mathematics\\
China University  of Mining and Technology\\
Beijing 100083, P R China} \maketitle

\begin{abstract}
By means of certain limit
technique, two kinds of generalized Darboux transformations are constructed  for the derivative
nonlinear Sch\"odinger equation (DNLS). These transformations are shown to lead to two solution formulas
for DNLS in terms of determinants. As applications, several
different types of high order solutions are calculated for this equation.

\end{abstract}
 \maketitle
\section{Introduction}
The derivative nonlinear Sch\"odinger equation (DNLS) \cite{M,MW}
\begin{equation}\label{dnlss}
    {\rm i}u_t+u_{xx}+{\rm i}(|u|^2u)_x=0,
\end{equation}
has many physical applications, especially in space plasma physics
and nonlinear optics. It well describes small-amplitude nonlinear
Alfv\'en waves in a low-$\beta$ plasma, propagating strictly
parallel or at a small angle to the ambient magnetic field. It was
shown that the DNLS also models large-amplitude
magnetohydrodynamic (MHD) waves in a high-$\beta$ plasma propagating
at an arbitrary angle to the ambient magnetic field. In nonlinear
optics, the modified nonlinear Schr\"odinger equation \cite{CL},
 which is gauge equivalent to DNLS, arises in the theory of
ultrashort femtosecond nonlinear pulses in optical fibres, when the
spectral width of the pulses becomes comparable with the carrier
frequency and the effect of self-steepening of the pulse should be
taken into account.

High order solitons describe the interaction between N solitons of
equal amplitude but having a particular chirp \cite{GS}. In the
terminology of inverse scattering transformation (IST), they
correspond to multiple-pole solitons. In the case of the Korteweg-de
Vries equation (KdV), the poles must be simple, that is the reason
why  high order nonsingular solitons do not exist. Indeed we could
obtain multiple-pole solutions by Darboux transformation (DT), such
as positon solutions \cite{Matveev}. The high order solitons for
nonlinear Schr\"odinger equation had been studied by many authors
\cite{GS,VY,ZS}. To the best of our knowledge, the high order
solitons of DNLS have never been reported. The aim of the present
paper is to show that such solutions may be obtained by generalized
Darboux transformations (gDT).

Recently, the rogue wave phenomenon \cite{KP}, which ``appears from
nowhere and disappears without a trace", has been a subject of
extensive study. Those waves, also known as freak, monster or giant
waves, is characterized with large amplitudes and  often appear on
the sea surface. One of the possible ways to explain the rogue waves
is the rogue wave solution and modulation instability and there are
a series of works done by Akhmediev's group \cite{AAS,AAS1,AKA}.
Different approaches have been proposed to construct the generalized
rogue wave solutions of nonlinear Sch\"odinger equation (NLS), for
example, the algebro-geometric method is adopted by Matveev et al
\cite{DPKM,DM}, Ohta and Yang work in the framework of Hirota
bilinear method while the present authors use the gDT as a tool
\cite{GLL}.  In the IST terminology, the high rogue wave corresponds
to multiple-pole solution at the branch spectral parameter in the
non-vanishing background \cite{KPR}. The first order rogue wave for
DNLS was obtained by Xu and coworkers  recently \cite{XHW}. However,
the high order rogue wave had never been studied. We will tackle
this problem by constructing generalized Darboux transformations.

The Darboux transformation \cite{DL,Mat}, which does not need to do the
inverse spectral analysis, provides a direct way to solve the Lax
pair equations algebraically. However, there is a defect that
classical DT cannot be iterated at the same spectral parameter. This
defect makes it impossible to construct the high order rogue wave solutions by
DT directly. Thus we must modify the DT method. In this work, we extend the
DT by the limit technique, so that it can be iterated at the same
spectral parameter. The modified transformation is referred as generalized Darboux
transformation (gDT).

The inverse scattering method was used to study DNLS with vanishing
background (VBC) and non-vanishing background (NVBC)
\cite{CH,CL,CLL,KI,L}. The $N$-bright soliton formula for DNLS was
established by Nakamura and Chen by the Hirota bilinear method
\cite{NC} and the DT for DNLS was constructed by Imai \cite{I} and
Steudel \cite{S} (see also \cite{XHW}).
 One key aim of this work is the construction of the gDT and based on it, the high order soliton solutions and rogue wave
are obtained. In addition to above two kinds of solutions, new $N$-solitons and high order rational solutions are also found.

The organization of this paper is as follows. In section 2, we
provide a rigorous proof for elementary DT for Kaup-Newell (KN)
system. Furthermore, based on the elementary DT, we construct for KN
system the binary DT, which is referred as the DT-II while the
elementary DT is referred as DT-I. We also iterate these DT and work
out the $N$-fold DT's both for DT-I and DT-II, and consider two
different kinds of reductions of the DT of the KN system to the
DNLS. In section 3, the generalized DT-I and DT-II are constructed
in detail by the limit technique. In section 4, we consider the
applications of the generalized DT (gDT) and calculate various high
order solutions, which include high order bright solitons with the
VBC, and high order rogue wave solutions. Final section concludes
the paper and offers some discussions.

\section{Darboux transformation for DNLS}

Let us start with the following system --- Kaup-Newell system \cite{KN}
\begin{equation}\label{KN}
    \left.
      \begin{array}{rl}
      {\rm i}u_t+u_{xx}-{\rm i}(u^2v)_x=0,  \\
       -{\rm i}v_t+v_{xx}+{\rm i} (uv^2)_x=0,
      \end{array}
    \right.
\end{equation}
which may be written as the compatibility condition
\begin{equation}\label{com}
    U_t-V_x+[U,V]=0,
\end{equation}
of the linear system or Lax pair \cite{Lne}
\begin{subequations}\label{lin}
    \begin{eqnarray}
      \Phi_x &=& U\Phi,\\
      \Phi_t &=& V\Phi,
    \end{eqnarray}
\end{subequations}
where
\[
 U=[-\frac{{\rm i}}{\zeta^2}\sigma_3+\frac{1}{\zeta}Q],
\;\; V=-\frac{2{\rm
i}}{\zeta^4}\sigma_3+\frac{2}{\zeta^3}Q-\frac{{\rm
i}}{\zeta^2}Q^2\sigma_3+\frac{1}{\zeta}Q^3-\frac{{\rm
i}}{\zeta}Q_x\sigma_3,
\]
with
\begin{equation*}
    \sigma_3=\begin{pmatrix}
               1 & 0 \\
               0 & -1 \\
             \end{pmatrix},\quad Q=\begin{pmatrix}
                                     0 & u \\
                                     v & 0 \\
                                   \end{pmatrix}.
\end{equation*}
These equations \eqref{KN} are reduced to the DNLS
\eqref{dnlss} for $v=-u^*$. For convenience, we introduce the
following conjugate linear system for \eqref{lin},
\begin{subequations}\label{conlin}
    \begin{eqnarray}
      -\Psi_x &=& \Psi U,\\
      -\Psi_t &=& \Psi V.
    \end{eqnarray}
\end{subequations}
\subsection{DT-I}

 In the following, we first
consider the DT of the unreduced linear system \eqref{lin}. Generally speaking, DT is a
special gauge transformation which keeps the form of Lax pair
equation invariant. The explicit steps for constructing DT in $1+1$
dimensional integrable system are as following: first, we consider the  the gauge transformation $$D[1]=\zeta
D_1+D_0,$$ where $D_1$ and $D_0$ are unknown matrices which do not
which depends on $\zeta$. Then imposing that $D[1]$ is a DT we
have
$$D[1]_x+D[1]U=U[1]D[1],\quad (\det(D[1]))_x=\mathrm{Tr}(U[1]-U)\det(D[1]),$$
where $U[1]$ represents transformed $U$ matrix.
 After some analysis,
we find the following elementary DT (eDT)  for \eqref{lin}:
\begin{equation}\label{eleDt}
    D[1]=\sigma_1\left(\zeta+\zeta_1-2\zeta_1 P_1\right),\quad
    P_1=\frac{\Phi_1\Phi_1^T\sigma_1}{\Phi_1^T\sigma_1\Phi_1},\quad
    \sigma_1=\begin{pmatrix}
               0 & 1 \\
               1 & 0 \\
             \end{pmatrix},
\end{equation}
where $\Phi_1=(\phi_1,\varphi_1)^T$ is a special solution for
linear system \eqref{lin} at $\zeta=\zeta_1$, and $\Phi_1^T\sigma_1$
is a special solution for conjugate linear system \eqref{conlin}
at $\zeta=-\zeta_1$.
 Here we point out that this DT for DNLS was first
derived by Imai \cite{I}. With the help of the DT, Steudel\cite{S} and Xu et al  \cite{XHW} calculated various solutions for DNLS. Next, we give a rigorous proof that the above transformation does qualify as a DT.
\begin{thm} With $\Phi_1$ and $D[1]$ defined above, $
\Phi[1]=D[1]\Phi
$
solves
\[
\Phi[1]_x=U[1]\Phi[1],\;\; \Phi[1]_t=V[1]\Phi[1]
\]
where
$$U[1]=-\frac{{\rm i}}{\zeta^2}\sigma_3+\frac{1}{\zeta}Q[1],\quad Q[1]=\sigma_1Q\sigma_1-2\zeta_1\sigma_1P_{1,x}\sigma_1,$$ and
$$V[1]=-\frac{2{\rm
i}}{\zeta^4}\sigma_3+\frac{2}{\zeta^3}Q[1]-\frac{{\rm
i}}{\zeta^2}Q[1]^2\sigma_3+\frac{1}{\zeta}Q[1]^3-\frac{{\rm
i}}{\zeta}Q[1]_x\sigma_3.
$$
Namely, the $D[1]$ qualifies as a Darboux matrix. Correspondingly, $D[1]^{-1}$ is a Darboux matrix for the  conjugate Lax system.
\end{thm}
\noindent
{\em Proof}:
To begin with, we notice
\begin{equation*}
    D[1]^{-1}=\frac{1}{\zeta+\zeta_1}(I+\frac{2\zeta_1}{\zeta-\zeta_1}P_1)\sigma_1.
\end{equation*}
What we need to do is to verify
\begin{eqnarray}
  U[1] &=& D[1]_{,x}D[1]^{-1}+D[1] U D[1]^{-1}, \label{x-part}\\
  V[1] &=& D[1]_{,t}D[1]^{-1}+D[1] V D[1]^{-1}.\label{t-part}
\end{eqnarray}

First we consider \eqref{x-part}. The residue
for function $F_1(\zeta)\equiv D[1]_{,x}D[1]^{-1}+D[1] U D[1]^{-1}-U[1]$ at
$\zeta=\zeta_1$ is
\begin{equation*}
    {\mathrm{Res}_{\zeta_1}}(F_1(\zeta))=2\zeta_1\sigma_1\left[-(I-P_1)P_{1,x}+(I-P_1)U(\zeta_1)P_1\right]\sigma_1=-2\zeta_1\sigma_1(I-P_1)\Phi_1\left[\frac{\sigma_1\Phi_1^T}{\sigma_1\Phi_1^T\Phi_1}\right]_x\sigma_1=0.
\end{equation*}
where we used the relation
$D[1]_{,x}D[1]^{-1}=-D[1](D[1]^{-1})_x$. Similarly, the residue of
function $F_1(\zeta)$ at $\zeta=-\zeta_1$ is
\begin{equation*}
    \mathrm{Res}_{-\zeta_1}(F_1(\zeta))=-2\zeta_1\sigma_1[-P_{1,x}(I-P_1)+P_1U(-\zeta_1)(I-P_1)]\sigma_1=2\zeta_1\sigma_1\left[\frac{\Phi_1}{\sigma_1\Phi_1^T\Phi_1}\right]_x\sigma_1\Phi_1^T(I-P_1)\sigma_1=0.
\end{equation*}
Due to
\[
U[1]=-\frac{1}{\zeta^2}\sigma_3+\frac{1}{\zeta}Q[1],
\]
 and
\begin{equation}\label{field}
Q[1]=\sigma_1Q\sigma_1-2\zeta_1\sigma_1P_{1,x}\sigma_1,
\end{equation}
 the function $F_1(\zeta)D[1]$ is equal to zero at $\zeta=0$.
Thus the function $F_1(\zeta)$ is analytic at $\zeta=0$. It is easy to see
that $F_1(\zeta)\rightarrow0$ at $\zeta\rightarrow\infty.$ Therefore the
equality \eqref{x-part} is valid.

Now we turn to the time evolution part
\eqref{t-part}. We introduce a matrix
$\widehat{V[1]}=-\frac{2{\rm
i}}{\zeta^4}\sigma_3+\frac{2}{\zeta^3}Q[1]+\frac{{\rm
i}}{\zeta^2}V_2+\frac{1}{\zeta}V_1$, so that
 $F_2(\zeta)\equiv D[1]_{,t}D[1]^{-1}+D[1] V
D[1]^{-1}-\widehat{V[1]}$. Proceeding similarly as above, it is found that $F_2(\zeta)$ is analytic at $\zeta=\pm\zeta_1,0$ and
tends to zero as $\zeta\rightarrow\infty$, thus $F_2(\zeta)\equiv 0$.

In the following, we show $\widehat{V[1]}=V[1]$. Because of the
compatibility condition $(D[1]\Phi)_{xt}=(D[1]\Phi)_{tx}$, we have
\begin{equation}\label{com[1]}
    U[1]_t-\widehat{V[1]}_x+\left[U[1],\widehat{V[1]}\right]=0.
\end{equation}
Identifying terms of $\mathrm{O}(\zeta)$ in \eqref{com[1]}, we have
\begin{equation}\label{pr1}
  [\sigma_3,V_2] = 0,
\end{equation}
\begin{equation}\label{pr2}
  [\sigma_{3},V_{1}]=[V_2,Q[1] ]+2Q[1]_x,
  \end{equation}\begin{equation}\label{pr3}
 [Q[1],V_1]= V_{2,x}.
\end{equation}
From \eqref{pr1}, we have $V_2^{\mbox{off}}=0$, where $V_2^{\mbox{off}}$ denotes the the off-diagonal part of $V_2$. Similarly, we have
\begin{equation}\label{v1}
V_1^{\mbox{off}}={\rm i}\sigma_3Q[1]_x-\frac{{\rm i}}{2}\sigma_3[Q[1],V_2]
\end{equation}
 through \eqref{pr2}.
Substituting \eqref{v1} into \eqref{pr3} and solving it yields
\begin{equation*}
    V_2=-{\rm i}Q[1]^2\sigma_3+f(t).
\end{equation*}
Letting $\zeta_1=0$, one can readily obtain
$V_2=-{\rm i}\sigma_1Q^2\sigma_1\sigma_3.$ Therefore we have
$f(t)=0$. Moreover, $V_1=Q[1]^3-{\rm i}Q[1]_x\sigma_3.$

Similar argument could show that $D[1]^{-1}$ does qualify as a Darboux matrix for the conjugate Lax system. Thus the proof is completed.
\begin{rem}
In addition to \eqref{field} there is a different representation for $Q[1]$
\begin{equation}\label{ren}
Q[1]=\sigma_1(I-2P_1)Q(I-2P_1)\sigma_1+\frac{2{\rm
i}}{\zeta_1}\sigma_3\sigma_1(I-2P_1)\sigma_1.
\end{equation}
which is appeared in papers \cite{S,XHW}. We will use \eqref{field} rather than \eqref{ren}, since the former is  more compact.
\end{rem}

To derive the $N$-fold DT for this elementary DT \eqref{eleDt}, which is referred as DT-I,
we rewrite it as
\begin{equation*}
    D[1]=\begin{pmatrix}
      - \zeta_1\frac{\varphi_1}{\phi_1} & \zeta \\
       \zeta & -\zeta_1\frac{\phi_1}{\varphi_1} \\
     \end{pmatrix}.
\end{equation*}
 Assuming $N$ different solutions
$\Phi_i=(\phi_1,\varphi_1)^T$ of  \eqref{lin} at $\zeta=\zeta_i$ are given, we may have \begin{prop}\cite{I,S,XHW}
The N-fold DT for DT-I can be represented as
\begin{equation}\label{DT-I}
    D_N=D[N]D[N-1]\cdots D[1]=\zeta^N\sigma_1^{N}+\sum_{k=0}^{N-1}\begin{pmatrix}
             \alpha_k & 0 \\
             0 & \beta_k \\
           \end{pmatrix}\sigma_1^k\zeta^{k},
\end{equation}
where $\alpha_i$  are determined by the following equations
$$\left\{
    \begin{array}{ll}
      \alpha_0\varphi_i+\alpha_1\zeta_i\phi_i+\cdots+\alpha_{2l-1}\zeta_i^{2l-1}\phi_i+\alpha_{2l}\zeta_i^{2l}\varphi_i=-\zeta_i^{2l+1}\varphi_i, & \hbox{when $N=2l+1$;}
\\[10pt]
      \alpha_0\varphi_i+\alpha_1\zeta_i\phi_i+\cdots+\alpha_{2l-2}\zeta_i^{2l-2}\varphi_i+\alpha_{2l-1}\zeta_i^{2l-1}\phi_i=-\zeta_i^{2l}\phi_i, & \hbox{when $N=2l$.}
    \end{array}
  \right.$$
$i=1,2,\cdots,N$. And
$\beta_i=\alpha_i(\varphi_j\leftrightarrow\phi_j), (j=1,2,\cdots,
N)$.

 The transformation between the fields is the following:
\begin{enumerate}
  \item
When $N=2l+1$
\begin{equation}\label{1-fields1}
    u[N]=-v-\left[\frac{\det(B)}{\det(A)}\right]_x,\quad
v[N]=-u+\left[\frac{\det(B(\varphi_j\leftrightarrow\phi_j))}{\det(A(\varphi_j\leftrightarrow\phi_j))}\right]_x,
\end{equation}
where $A=\left(A_1^T,A_2^T,\cdots,A_N^T\right)$,
$B=\left(B_1^T,B_2^T,\cdots,B_N^T\right)$,
$$A_i= (\varphi_i,\zeta_i\phi_i,\cdots,\zeta_i^{2l-1}\phi_i,\zeta_i^{2l}\varphi_i ),$$
$$B_i= (\varphi_i,\zeta_i\phi_i,\cdots,\zeta_i^{2l-1}\phi_i,\zeta_i^{2l+1}\phi_i ).$$
 \item
When $N=2l$
\begin{equation}\label{1-fields2}
    u[N]=u-\left[\frac{\det(D)}{\det(C)}\right]_x,\quad
v[N]=v+\left[\frac{\det(D(\varphi_j\leftrightarrow\phi_j))}{\det(C(\varphi_j\leftrightarrow\phi_j))}\right]_x,
\end{equation}
where $C=\left(C_1^T,C_2^T,\cdots,C_N^T\right)$,
$D=\left(D_1^T,D_2^T,\cdots,D_N^T\right)$,
$$C_i=(\varphi_i,\zeta_i\phi_i,\cdots,\zeta_i^{2l-2}\varphi_i,\zeta_i^{2l-1}\phi_i),$$
$$D_i=(\varphi_i,\zeta_i\phi_i,\cdots,\zeta_i^{2l-2}\varphi_i,\zeta_i^{2l}\varphi_i).$$
\end{enumerate}
\end{prop}

\subsection{DT-II}

In this subsection, we will show that the so-called
dressing-B\"{a}cklund transformation \cite{Lne,W},  denoted by DT-II
in the present paper, may be constructed from above  DT-I. For
convenience, we rewrite $D[1]$ and $D[1]^{-1}$ as following
\begin{equation}\label{t1w}
 D[1]=\begin{pmatrix}
      - \zeta_1\frac{\varphi_1}{\phi_1} & \zeta \\
       \zeta & -\zeta_1\frac{\phi_1}{\varphi_1} \\
     \end{pmatrix},\quad D[1]^{-1}=\frac{1}{\zeta^2-\zeta_1^2}\begin{pmatrix}
       \zeta_1\frac{\phi_1}{\varphi_1} & \zeta \\
       \zeta & \zeta_1\frac{\varphi_1}{\phi_1} \\
     \end{pmatrix}.
\end{equation}
Suppose another solution $\Psi_1=(\chi_1,\psi_1)$ for the
conjugate system \eqref{conlin} at $\zeta=\xi_1$ is given, then
$\Psi_1[1]=\Psi_1D[1]^{-1}|_{\zeta=\xi_1}$ is a new solution for
conjugate system $(\Psi[1],U[1],V[1])$ at $\zeta=\xi_1$. It is easy to see that $\sigma_1\Psi_1[1]^T$ is a special
solution for Lax pair $(\Phi[1],U[1],V[1])$ at $\zeta=-\xi_1$.
Therefore, we could construct the second step DT $D[2]$ by the seed
solution $\sigma_1\Psi_1[1]^T.$ By direct calculations, removing the
factor $\zeta^2-\xi_1^2$, we have the DT-II
\begin{equation}\label{t1t2}
    T[1]=I+\frac{A}{\zeta-\xi_1}-\frac{\sigma_3A\sigma_3}{\zeta+\xi_1},\quad
    A=\frac{\xi_1^2-\zeta_1^2}{2}\begin{pmatrix}
                                     \alpha & 0 \\
                                     0 & \beta \\
                                   \end{pmatrix}\Phi_1\Psi_1,
\end{equation}
where
\begin{equation*}
    \alpha^{-1}=\Psi_1\begin{pmatrix}
                   \xi_1 & 0 \\
                   0 & \zeta_1 \\
                 \end{pmatrix}\Phi_1,\quad\beta^{-1}=\Psi_1\begin{pmatrix}
                   \zeta_1 & 0 \\
                   0 & \xi_1 \\
                 \end{pmatrix}\Phi_1.
\end{equation*}
Furthermore, we have
\begin{equation}\label{t1t2-1}
    T[1]^{-1}=I+\frac{B}{\zeta-\zeta_1}-\frac{\sigma_3B\sigma_3}{\zeta+\zeta_1},\quad
    B=\frac{\zeta_1^2-\xi_1^2}{2}\Phi_1\Psi_1\begin{pmatrix}
                                     \beta & 0 \\
                                     0 & \alpha \\
                                   \end{pmatrix}.
\end{equation}
The transformation between $Q$ and $Q[2]$ is
\begin{equation}\label{fileds}
    Q[2]=Q+\left[A-\sigma_3A\sigma_3\right]_x.
\end{equation}
Above discussion indicates that the DT-II $T[1]$ is indeed a two-fold DT for $D[1]$ in the case of
DNLS. We remark that in the case of two component DNLS the analogy of DT-II exists \cite{LL} while the corresponding DT-I has not been constructed.

In what follows, we consider the iteration for the DT-II $T[1]$.
Assume we have $N$ distinct solutions
$\Phi_i(\mu_i)=(\phi_i,\varphi_i)^T$ of \eqref{lin} at $\zeta=\mu_i$
and $N$ distinct solutions $\Psi_i(\nu_i)=(\chi_i,\psi_i)$ of
\eqref{conlin} at $\zeta=\nu_i$. Similar to DT-I, we work with
DT-II $T[1]$  and  have the following proposition
\begin{prop}
The N-fold DT for the DT-II could be written as the following form
\begin{equation}\label{n-fold}
    T_N=T[N]T[N-1]\cdots T[1]=I+\sum_{i=1}^{N}\left(\frac{C_i}{\zeta-\nu_i}-\frac{\sigma_3C_i\sigma_3}{\zeta+\nu_i}\right)
\end{equation}
and
\begin{equation}\label{n-fold1}
    T_N^{-1}=T[1]^{-1}T[2]^{-1}\cdots T[N]^{-1}=I+\sum_{i=1}^{N}\left(\frac{D_i}{\zeta-\mu_i}-\frac{\sigma_3D_i\sigma_3}{\zeta+\mu_i}\right)
\end{equation}
\end{prop}
\noindent
{\em Proof}: We calculate the residues for both sides of \eqref{n-fold}
\begin{eqnarray*}
  \mathrm{Res}|_{\zeta=\nu_i}(T_N) &=&(I+\frac{A_{N}}{\nu_i-\nu_N}-\frac{\sigma_3A_{N}\sigma_3}{\nu_i+\nu_N})\cdots A_i\cdots(I+\frac{A_{1}}{\nu_i-\nu_1}-\frac{\sigma_3A_{1}\sigma_3}{\nu_i+\nu_1}), \\
  \mathrm{Res}|_{\zeta=-\nu_i}(T_N) &=&-(I+\frac{A_{N}}{-\nu_i-\nu_N}-\frac{\sigma_3A_{N}\sigma_3}{-\nu_i+\nu_N})\cdots
\sigma_3A_i\sigma_3\cdots(I+\frac{A_{1}}{-\nu_i-\nu_1}-\frac{\sigma_3A_{1}\sigma_3}{-\nu_i+\nu_1}).
\end{eqnarray*}
Because of $\mathrm{Res}|_{\zeta=\nu_i}(T_N)=-\sigma_3\mathrm{Res}|_{\zeta=-\nu_i}(T_N)\sigma_3$,  equation \eqref{n-fold} is valid. Similarly, \eqref{n-fold1} can be proved.

The N-fold DT-II
$T_N$ allows us to find the transformations between
 the fields $u[0]$, $v[0]$ and $u[N]$, $v[N]$, which are given below
\begin{thm}
The N-fold DT-II $T_N$ induces the following transformations for the fields
\begin{equation}\label{filedsn}
u[N]=u[0]-2\left(\frac{\det M_1 }{\det M}\right)_x,\quad
v[N]=v[0]+2\left(\frac{\det N_1 }{\det N}\right)_x,
\end{equation}
where $M=(M_{ij})_{N\times N},$ $N=(N_{ij})_{N\times N}$,
$M_{ij}=\frac{\Psi_i\sigma_3\Phi_j}{\mu_j+\nu_i}-\frac{\Psi_i\Phi_j}{\mu_j-\nu_i},$
$N_{ij}=-\left[\frac{\Psi_i\sigma_3\Phi_j}{\mu_j+\nu_i}+\frac{\Psi_i\Phi_j}{\mu_j-\nu_i}\right],$
\begin{equation*}
    M_1=\begin{pmatrix}
          M_{11} & M_{12} & \cdots & M_{1N} & \psi_1 \\
          M_{21} & M_{22} & \cdots & M_{2N} & \psi_2 \\
          \vdots & \vdots&\ddots  & \vdots & \vdots \\
          M_{N1} & M_{N2} & \cdots & M_{NN} & \psi_N \\
          \phi_1 & \phi_2 & \cdots  & \phi_N & 0  \\
        \end{pmatrix},\quad N_1=\begin{pmatrix}
          N_{11} & N_{12} & \cdots & N_{1N} & \chi_1 \\
          N_{21} & N_{22} & \cdots & N_{2N} & \chi_2 \\
          \vdots & \vdots&\ddots  & \vdots & \vdots \\
          N_{N1} & N_{N2} & \cdots & N_{NN} & \chi_N \\
          \varphi_1 & \varphi_2 & \cdots  & \varphi_N & 0  \\
        \end{pmatrix}.
\end{equation*}
\end{thm}
\noindent
{\bf Proof}: Since $T_N$ \eqref{n-fold} is the N-fold DT  of \eqref{lin}, we have
\begin{equation*}
    T_{N,x}+T_NU=U[N]T_N.
\end{equation*}
It follows that
$$Q[N]=Q+\sum_{i=1}^{N}[C_i-\sigma_3C_i\sigma_3]_x.$$

Thus, we need to calculate the explicit forms for $C_i$. Proposition 2 implies $C_i=\mathrm{Res}|_{\zeta=\mu_i}(T_N)$, it
implies  that $C_i$'s are the matrices of rank one. Thus we may assume
$C_i=|x_i\rangle\langle y_i|$. Similarly we may set $D_i=|w_i\rangle\langle v_i|$.

On the one hand, because of $T_NT_N^{-1}=I$, we have
\begin{equation}\label{relation1}
    \langle y_l|T_N^{-1}|_{\zeta=\nu_l}=0,
\end{equation}
where the fact that the residue $T_NT_N^{-1}$ at $\zeta=\nu_l$ 
equals to zero is taken account of. 
On the other hand, we have
$$\Psi_lT_N^{-1}|_{\zeta=\nu_l}=0.$$
Noticing that the rank of $T_N^{-1}|_{\zeta=\nu_i}$ equals to 1, we
may obtain
$$\langle y_l|=\Psi_l.$$
Now substituting  $\langle y_l|$ into \eqref{relation1} leads to
\begin{equation}\label{relation2}
    \Phi_l+\sum_{i=1}^{N}\left(\frac{|x_i\rangle
    \Psi_i\Phi_l}{\mu_l-\nu_i}-\frac{\sigma_3|x_i\rangle\Psi_i\sigma_3\Phi_l}{\mu_l+\nu_i}\right)=0,
    \;\; (l=1,2,...,N).
\end{equation}
Solving \eqref{relation2} gives us
\begin{eqnarray*}
  \left[|x_1\rangle,|x_2\rangle,\cdots,|x_N\rangle \right]_1 &=& \left[\phi_1,\phi_2,\cdots,\phi_N \right]M^{-1}, \\
  \left[|x_1\rangle,|x_2\rangle,\cdots,|x_N\rangle \right]_2 &=& \left[\varphi_1,\varphi_2,\cdots,\varphi_N
  \right]N^{-1},
\end{eqnarray*}
where subscript $_1$ and $_2$ stand the first and second rows
respectively. Finally, the relations between the fields can be
represented as
\begin{eqnarray*}
  u[N] &=& u[0]+2\left[(\phi_1,\phi_2,\cdots,\phi_N)M^{-1}\begin{pmatrix}
                                                                  \psi_1 \\
                                                                  \psi_2 \\
                                                                  \vdots \\
                                                                  \psi_N \\
                                                                \end{pmatrix}
  \right]_x=u[0]-2\left(\frac{\det M_1 }{\det M}\right)_x, \\
  v[N] &=& v[0]-2\left[(\varphi_1,\varphi_2,\cdots,\varphi_N)N^{-1}\begin{pmatrix}
                                                                  \chi_1 \\
                                                                  \chi_2 \\
                                                                  \vdots \\
                                                                  \chi_N \\
                                                                \end{pmatrix}
  \right]_x=v[0]+2\left(\frac{\det N_1 }{\det N}\right)_x.
\end{eqnarray*}
This completes the proof.

\subsection{Reduction}
So far we have been working with the DTs for the general Lax problem \eqref{lin} and certain solution formulae have been  given for the system \eqref{KN}. However our main task is to construct solutions for DNLS \eqref{dnlss}, therefore we have to consider reduction problem. It is easy to see that two reductions $v=u^*$ and $v=-u^*$ are simply related \cite{S}, so we may consider either of them.  For the DT-I, let us assume $v=-u^*$ or $Q^{\dag}=-Q$,
where $^{\dag}$ denotes the complex conjugation and transpose. To implement the reduction, we need to choose the seed solutions properly. Indeed, assuming
\begin{equation}\label{reduction1}
\zeta_1\in{\rm i}\mathds{R},\quad \text{and}\quad
\varphi_1=\phi_1^*,
\end{equation}
then $Q[1]$, defined by \eqref{field}, satisfies the reduction relation $Q[1]^{\dag}=-Q[1]$.
The DT
\eqref{eleDt} with the reduction condition \eqref{reduction1} may be employed to construct  bright or dark solitons of DNLS with the non-vanishing background.

Let us now turn to the reduction of the DT-II. Assuming $Q=Q^{\dag}$ and
\begin{equation}\label{reduction2}
    \xi_1=\zeta_1^*, \quad\text{and}\quad
(\chi_1,\psi_1)=(\phi_1^*,\varphi_1^*),
\end{equation}
\eqref{fileds} yields $Q[2]=Q[2]^{\dag}$.

To iterate the DT-I and DT-II, we must verify that they keep
the reduction conditions \eqref{reduction1} and \eqref{reduction2}. The latter
 merely depends on the symmetry of
equations \eqref{lin}, thus it holds automatically. For the
former \eqref{reduction1} we claim that
\begin{prop}
Both  DT $D[1]$ and $T[1]$ keep the reduction condition \eqref{reduction1}
invariant.
\end{prop}
\noindent
{\em  Proof}: Direct calculations.

Due to above analysis, both DT-I and DT-II may be reduced to find
solutions for DNLS. However, the DT-I under \eqref{reduction1} is
conveniently used  to construct the  N-dark or bright soliton
solutions of DNLS with NVBC, while DT-II with \eqref{reduction2} may
be properly adopted to represent the N-bright solitons and
N-breathers of DNLS.

\section{Generalized Darboux transformations}
In this section, we construct the corresponding generalized Darboux
transformations (gDT) associated with $D[1]$ and $T[1]$. We will
follow the approach proposed for the nonlinear Sch\"{o}rdinger
equation in \cite{GLL}. Indeed, while both DT-I and DT-II considered
above are degenerate at $\zeta=\zeta_1$ in the sense that
$D[1]|_{\zeta=\zeta_1}\Phi_1=T[1]|_{\zeta=\zeta_1}\Phi_1=0$, we may
%
work with
$$\Phi_1^{[1]}=\lim_{\epsilon\rightarrow0}\frac{(D[1]\Phi_1)|_{\zeta=\zeta_1+\epsilon}}{\epsilon}$$
or
$$\Phi_1^{[1]}=\lim_{\epsilon\rightarrow0}\frac{(T[1]\Phi_1)|_{\zeta=\zeta_1+\epsilon}}{\epsilon}$$
which serves the seed solution for doing next step transformation.
%

\subsection{gDT-I}
To construct the gDT associated with DT-I, we assume that $n$ solutions $(\varphi_i,\phi_i)^T$ are given for the Lax pair at $\zeta=\zeta_i$ $(i=1, \cdots, n)$.
First, we have the
elementary DT
\begin{equation*}
    D_1^{[0]}=\begin{pmatrix}
          -\zeta_1\frac{\phi_1}{\varphi_1} & \zeta \\
          \zeta & -\zeta_1\frac{\varphi_1}{\phi_1} \\
        \end{pmatrix}.
\end{equation*}
As observed above, by virtue of the limit process, we find that
\begin{equation*}
    \begin{pmatrix}
      \varphi_1^{[1]} \\
      \phi_1^{[1]} \\
    \end{pmatrix}=\lim_{\epsilon\rightarrow 0}\frac{D_1^{[0]}\big|_{\zeta=\zeta_1}+\epsilon\sigma_1}{\epsilon}\begin{pmatrix}
                                                                                                      \varphi_1(\zeta_1+\epsilon) \\
                                                                                                      \phi_1(\zeta_1+\epsilon) \\
                                                                                                    \end{pmatrix}=D_1^{[0]}\big|_{\zeta=\zeta_1}
\frac{\mathrm{d}}{\mathrm{d}\zeta}\begin{pmatrix}
                                    \varphi_1(\zeta) \\
                                    \phi_1(\zeta) \\
                                  \end{pmatrix}_{\zeta=\zeta_1}+\sigma_1\begin{pmatrix}
                                    \varphi_1(\zeta_1) \\
                                    \phi_1(\zeta_1) \\
                                  \end{pmatrix}
\end{equation*} is a non-trivial solution for Lax pair \eqref{lin} with $u=u[1]$ and $v=v[1]$ at $\zeta=\zeta_1$, which may lead to the next step transformation
\begin{equation*}
    D_1^{[1]}=\begin{pmatrix}
          -\zeta_1\frac{\phi_1^{[1]}}{\varphi_1^{[1]}} & \zeta \\
          \zeta & -\zeta_1\frac{\varphi_1^{[1]}}{\phi_1^{[1]}} \\
        \end{pmatrix}.
\end{equation*}
This process may be continued and we have the following theorem.
\begin{thm}
Let $(\varphi_i,\phi_i)^T$ be the solutions of Lax pair at $\zeta=\zeta_i$ $(i=1, \cdots,  n)$ and assume DT-I possesses $m_i$ order zeros at $\zeta=\zeta_i$. Then we have the following gDT-I:
\begin{equation}\label{gdt1}
    D_N=D_n^{[m_n-1]}\cdots D_n^{[1]}D_n^{[0]}\cdots D_1^{[m_1-1]}\cdots D_1^{[1]}
    D_1^{[0]}
\end{equation}
where $$N=\sum^{n}_{i=1}m_i,$$ and
\begin{equation*}
    D_i^{[j]}=\begin{pmatrix}
                -\zeta_i\frac{\phi_i^{[j-1]}}{\varphi_i^{[j-1]}} & \zeta \\
                \zeta & -\zeta_i\frac{\varphi_i^{[j-1]}}{\phi_i^{[j-1]}} \\
              \end{pmatrix},
\end{equation*}
\begin{equation}\label{gdt1t}
    \begin{pmatrix}
      \varphi_i^{[j-1]}\\
      \phi_i^{[j-1]} \\
    \end{pmatrix}=\sum_{l=1}^{j-1}\frac{\Omega_{j-1-l}}{l!}\frac{\mathrm{d}}{d\zeta^l}\begin{pmatrix}
      \varphi_i\\
      \phi_i \\
    \end{pmatrix}\big|_{\zeta=\zeta_i},\quad \begin{pmatrix}
      \varphi_i^{[0]}\\
      \phi_i^{[0]} \\
    \end{pmatrix}=\begin{pmatrix}
      \varphi_i\\
      \phi_i \\
    \end{pmatrix},
\end{equation}
and
\begin{equation*}
   \Omega_l=\sum_{\sum \delta_i^k=l}M_{i}^{[j-2]}\cdots M_{i}^{[0]}\cdots M_{1}^{[m_1-1]}\cdots M_{1}^{[0]}, \quad M_i^{[k]}=\left\{
       \begin{array}{ll}
         \sigma_1, & \hbox{if  $\delta_i^{k}=1$;} \\[10pt]
         D_i^{[k]}\big|_{\zeta=\zeta_i}, & \hbox{if $\delta_i^{k}=0$.}
       \end{array}
     \right.
\end{equation*}

\end{thm}
\noindent
{\em Proof}: To construct the gDT-I, we start with the
eDT
\begin{equation*}
    D_1^{[0]}=\begin{pmatrix}
          -\zeta_1\frac{\phi_1}{\varphi_1} & \zeta \\
          \zeta & -\zeta_1\frac{\varphi_1}{\phi_1} \\
        \end{pmatrix}.
\end{equation*}
By means of the nontrivial solutions
$(\varphi_1[1],\phi_1[1])$, we may do the next step of transformation $D_1^{[1]}$.
Taking account of the given seeds $(\varphi_i,\phi_i)^T$, we perform the following limit
\begin{eqnarray*}
    \begin{pmatrix}
      \varphi_i^{[j]} \\
      \phi_i^{[j]} \\
    \end{pmatrix}&=&\lim_{\epsilon\rightarrow 0}\frac{\left[D_i^{[j-1]}\cdots D_i^{[1]}D_i^{[0]}\cdots D_1^{[m_1-1]}\cdots D_1^{[1]} D_1^{[0]}\right]\Big|_{\zeta=\zeta_i+\epsilon}}{\epsilon^{j}}\begin{pmatrix}
      \varphi_i(\zeta_i+\epsilon) \\
      \phi_i(\zeta_i+\epsilon) \\
    \end{pmatrix},
\end{eqnarray*}
which yields the formulae presented in above theorem. This completes the proof.

To have a compact determinantal representation for the gDT-I, we may take the limit
 directly on the N-fold DT-I \eqref{DT-I}. It follows from
\eqref{1-fields1} and \eqref{1-fields2} that the transformations between
the fields are:
\begin{enumerate}
  \item When $N=2l+1$
\begin{equation}\label{1-fields11}
    u[N]=-v[0]-\left[\frac{\det(B)}{\det(A)}\right]_x,\quad
v[N]=-u[0]+\left[\frac{\det(B(\varphi_j\leftrightarrow\phi_j))}{\det(A(\varphi_j\leftrightarrow\phi_j))}\right]_x,
\end{equation}
where
$$A=\left(A_1^T,\frac{\mathrm{d}}{\mathrm{d}\zeta}A_1^T,\cdots,\frac{d^{m_1-1}}{(m_1-1)!\mathrm{d}\zeta^{m_1-1}}A_1^T,
\cdots,A_n^T,\frac{\mathrm{d}}{\mathrm{d}\zeta}A_n^T,\cdots,\frac{d^{m_n-1}}{(m_n-1)!\mathrm{d}\zeta^{m_n-1}}A_n^T\right),$$
$$B=\left(B_1^T,\frac{\mathrm{d}}{\mathrm{d}\zeta}B_1^T,\cdots,\frac{d^{m_1-1}}{(m_1-1)!\mathrm{d}\zeta^{m_1-1}}B_1^T,
\cdots,B_n^T,\frac{\mathrm{d}}{\mathrm{d}\zeta}B_n^T,\cdots,\frac{d^{m_n-1}}{(m_n-1)!\mathrm{d}\zeta^{m_n-1}}B_n^T\right).$$
  and $A_i$, $B_i$ are the same as \eqref{1-fields1}.
\item
When $N=2l$\begin{equation}\label{1-fields21}
    u[N]=u[0]-\left[\frac{\det(D)}{\det(C)}\right]_x,\quad
v[N]=v[0]+\left[\frac{\det(D(\varphi_j\leftrightarrow\phi_j))}{\det(C(\varphi_j\leftrightarrow\phi_j))}\right]_x,
\end{equation}
where
$$C=\left(C_1^T,\frac{\mathrm{d}}{\mathrm{d}\zeta}C_1^T,\cdots,\frac{d^{m_n-1}}{(m_n-1)!\mathrm{d}\zeta^{m_n-1}}C_1^T,
\cdots,C_n^T,\frac{\mathrm{d}}{\mathrm{d}\zeta}C_n^T,\cdots,\frac{d^{m_n-1}}{(m_n-1)!\mathrm{d}\zeta^{m_n-1}}C_n^T\right),$$
$$D=\left(D_1^T,\frac{\mathrm{d}}{\mathrm{d}\zeta}D_1^T,\cdots,\frac{d^{m_n-1}}{(m_n-1)!\mathrm{d}\zeta^{m_n-1}}D_1^T,
\cdots,D_n^T,\frac{\mathrm{d}}{\mathrm{d}\zeta}D_n^T,\cdots,\frac{d^{m_n-1}}{(m_n-1)!\mathrm{d}\zeta^{m_n-1}}D_n^T\right).$$
and $C_i$, $D_i$ are the same as \eqref{1-fields2}.
\end{enumerate}

Thus, we complete the construction of gDT-I for \eqref{lin}, which
could be considered as a generalization for DT studied in
\cite{I,XHW,S}.

\subsection{gDT-II}
In this subsection, we consider the generalization for DT-II. To this end, we assume that $n$
solutions $\Phi_i(\zeta=\mu)=(\phi_i,\varphi_i)^T$ are given for the
Lax pair at $\mu=\mu_i$ and $n$ solutions
$\Psi_i(\zeta=\nu)=(\chi_i,\psi_i)$ are given for the conjugate Lax
pair at $\nu=\nu_i$ $(i=1, \cdots, n)$.


\begin{thm}
Let $\Phi_i$ be the solutions of Lax pair at $\zeta=\mu_i$ and
$\Psi_i$ be the solutions of conjugate Lax pair at $\zeta=\nu_i$
$(i=1, \cdots,  n)$,$$\sum_{i=1}^r m_i=N,$$
 assume DT-II possesses $m_i$ order zeros at
$\zeta=\pm\mu_i$ and inverse of DT-II possesses $m_i$ order zeros at
$\zeta=\pm\nu_i$. Then we have the following gDT-II
\begin{equation}\label{gdtii}
     T_N=T_n^{[m_i-1]}\cdots T_n^{[0]}\cdots T_1^{[m_1-1]}\cdots T_1^{[0]},\quad T_N^{-1}=T_1^{[0]-1}\cdots
     T_1^{[m_1-1]-1}\cdots T_n^{[0]-1}\cdots T_n^{[m_i-1]-1}
\end{equation}
where
\begin{equation*}
    T_i^{[j]}=I+\frac{A_i^{[j]}}{\zeta-\nu_i}-\frac{\sigma_3 A_i^{[j]}
    \sigma_3}{\zeta+\nu_i},\quad T_i^{[j]-1}=I+\frac{B_i^{[j]}}{\zeta-\mu_i}-\frac{\sigma_3
    B_i^{[j]}
    \sigma_3}{\zeta+\mu_i},
\end{equation*}
\begin{eqnarray*}
  A_i^{[j]} =\frac{\nu_i^2-\mu_i^2}{2}\begin{pmatrix}
                                     \alpha_i^{[j]} & 0 \\
                                     0 & \beta_i^{[j]} \\
                                   \end{pmatrix}\Phi_i^{[j]}\Psi_i^{[j]}, &&
  B_i^{[j]}=\frac{\mu_i^2-\nu_i^2}{2}\Phi_i^{[j]}\Psi_i^{[j]}\begin{pmatrix}
                                     \beta_i^{[j]} & 0 \\
                                     0 & \alpha_i^{[j]} \\
                                   \end{pmatrix},
\end{eqnarray*}
\begin{eqnarray*}
    \alpha_i^{[j]-1}=\Psi_i^{[j]}\begin{pmatrix}
                   \nu_i & 0 \\
                   0 & \mu_i \\
                 \end{pmatrix}\Phi_i^{[j]},&&\beta_i^{[j]-1}=\Psi_i^{[j]}\begin{pmatrix}
                   \mu_i & 0 \\
                   0 & \nu_i \\
                 \end{pmatrix}\Phi_i^{[j]},
\end{eqnarray*}
and
\begin{equation*}
    \Phi_i^{[j]}=\displaystyle{\sum_{l=0}^{j}}\frac{\Omega_l}{(j-l)!}\frac{\mathrm{d}^{j-l}}{\mathrm{d}\mu^{j-l}}\Phi_i|_{\mu=\mu_i},\quad  \Omega_l=\sum^{\sum\delta_i^j=l}M_{i}^{[j-1]}\cdots M_{i}^{[0]}\cdots M_{1}^{[m_1-1]}\cdots M_{1}^{[0]},
\end{equation*}
\begin{equation*}
    \Psi_i^{[j]}=\displaystyle{\sum_{l=0}^{j}}\frac{\mathrm{d}^{j-l}}{\mathrm{d}\mu^{j-l}}\Psi_i|_{\nu=\nu_i}\frac{\Lambda_l}{(j-l)!},\quad \Lambda_l=\sum^{\sum\delta_i^{j}=l}N_{i}^{[j-1]}\cdots N_{i}^{[0]}\cdots N_{1}^{[m_1-1]}\cdots N_{1}^{[0]},
\end{equation*}
\begin{equation*}
    M_i^{[j]}=\left\{
       \begin{array}{ll}
         \frac{1}{\mu_i^2-\nu_i^2}, & \hbox{if $\delta_i^{j}=2$;} \\[10pt]
         \frac{2\mu_i+A_i^{[j]}-\sigma_3A_i^{[j]}\sigma_3}{\mu_i^2-\nu_i^2}, & \hbox{if $\delta_i^{j}=1$;} \\[10pt]
         T_i^{[j]}\big|_{\zeta=\mu_i}, & \hbox{if $\delta_i^{j}=0$.}
       \end{array}
     \right.
,\quad
   N_i^{[j]}=\left\{
       \begin{array}{ll}
         \frac{1}{\nu_i^2-\mu_i^2}, & \hbox{if $\delta_i^{j}=2$;}
         \\[10pt]
         \frac{2\nu_i+B_i^{[j]}-\sigma_3B_i^{[j]}\sigma_3}{\nu_i^2-\mu_i^2}, & \hbox{if $\delta_i^{j}=1$;} \\[10pt]
         T_i^{[j]-1}\big|_{\zeta=\nu_i}, & \hbox{if $\delta_i^{j}=0$.}
       \end{array}
     \right.
\end{equation*}
\end{thm}
\noindent {\em Proof}: Noting that the DT-II is given \eqref{t1t2} and using
the limit technique, we could obtain special solutions for new Lax
pair \eqref{lin} $(\Phi[1],U[1],V[1])$ at $\zeta=\mu_1$ and
conjugate Lax pair \eqref{conlin} $(\Psi[1],U[1],V[1])$ at
$\zeta=\nu_1$, i.e.
\begin{eqnarray*}
  \Phi_1^{[1]} &=&\lim_{\delta\rightarrow 0} \frac{T_1^{[0]}|_{\zeta=\mu_1+\delta}\Phi_1(\mu_1+\delta)}{\delta}=T_1^{[0]}|_{\zeta=\mu_1}\frac{\mathrm{d}}{\mathrm{d}\mu}\Phi_1|_{\mu=\mu_1}+S_1\Phi_1(\mu_1),\\
  \Psi_1^{[1]} &=&\lim_{\delta\rightarrow 0}
  \frac{\Psi_1(\nu_1+\delta)T_1^{[0]-1}|_{\zeta=\nu_1+\delta}}{\delta}=\frac{\mathrm{d}}{\mathrm{d}\nu}\Psi_1|_{\nu=\nu_1}T_1^{[0]-1}|_{\zeta=\nu_1}+\Psi_1(\nu_1)R_1,
\end{eqnarray*}
where
$$S_1=\frac{2\mu_1+A_1-\sigma_3A_1\sigma_3}{\mu_1^2-\nu_1^2},\quad
R_1=\frac{2\nu_1+B_1-\sigma_3B_1\sigma_3}{\nu_1^2-\mu_1^2}.
$$
Therefore, we could continue to construct the DT-II $T[2]$ for the
new system
$$T_1^{[1]}=I+\frac{A_1^{[1]}}{\zeta-\nu_1}-\frac{\sigma_3A_1^{[1]}\sigma_3}{\zeta+\nu_1},\quad T_1^{[1]-1}=I+\frac{B_1^{[1]}}{\zeta-\mu_1}-\frac{\sigma_3B_1^{[1]}\sigma_3}{\zeta+\mu_1}.$$
Generally, taking account of the given seeds $\Phi_i$ and $\Psi_i$,
we perform the following limit
\begin{eqnarray*}
  \Phi_i^{[j]} &=&\lim_{\delta\rightarrow 0} \frac{\left( T_i^{[j-1]}\cdots T_i^{[0]}\cdots T_1^{[m_1-1]}\cdots T_1^{[0]}\right)|_{\zeta=\mu_i+\delta}}{\delta^j}\Phi_i(\mu_i+\delta),\\
  \Psi_i^{[j]} &=&\lim_{\delta\rightarrow 0}
  \Psi_i(\nu_i+\delta)\frac{\left( T_1^{[0]-1}\cdots
     T_1^{[m_1-1]-1}\cdots T_i^{[0]-1}\cdots T_i^{[j-1]-1}\right)|_{\zeta=\nu_i+\delta}}{\delta^j},
\end{eqnarray*}
and mathematical
induction leads to gDT-II \eqref{gdtii}. This completes the proof.

Due to above proposition, the transformations between the fields are
\begin{eqnarray}
                                           u[N] &=& u[0]+ \sum_{i=1}^{n}\sum_{j=0}^{m_i-1}\left[A_i^{[j]}-\sigma_3A_i^{[j]}\sigma_3\right]_x,\label{gdtii-u}\\
                                           v[N] &=& v[0]+
                                           \sum_{i=1}^{n}\sum_{j=0}^{m_i-1}\left[B_i^{[j]}-\sigma_3B_i^{[j]}\sigma_3\right]_x.\label{gdtii-v}
                                         \end{eqnarray}
As before, the above formulas \eqref{gdtii-u} and
\eqref{gdtii-v} could be rewritten with the determinant form, i.e.
\begin{equation}\label{filedsn1}
u[N]=u[0]-2\left(\frac{P_1 }{P}\right)_x,\quad
v[N]=v[0]-2\left(\frac{Q_1 }{Q}\right)_x
\end{equation}
where
\begin{equation*}
    P_1=\begin{pmatrix}
          P^{[11]} & P^{[12]} & \cdots & P^{[1r]} & \widehat{\psi_1 }\\
          P^{[21]} & P^{[22]} & \cdots & P^{[2r]} & \widehat{\psi_2} \\
          \vdots & \vdots&\ddots  & \vdots & \vdots \\
          P^{[r1]} & P^{[r2]} & \cdots & P^{[rr]} &  \widehat{\psi_r} \\
          \widehat{\phi_1} & \widehat{\phi_2} & \cdots  & \widehat{\phi_r} & 0  \\
        \end{pmatrix},\;\; P=\begin{pmatrix}
          P^{[11]} & P^{[12]} & \cdots & P^{[1r]} \\
          P^{[21]} & P^{[22]} & \cdots & P^{[2r]}  \\
          \vdots & \vdots&\ddots  & \vdots  \\
          P^{[r1]} & P^{[r2]} & \cdots & P^{[rr]}  \\
        \end{pmatrix},
\end{equation*}
with
$$\widehat{\psi_i}=\begin{pmatrix}
                     \psi_i, &
                     \frac{\partial}{\partial \nu}\psi_i, &
                     \cdots, &
                     \frac{1}{(m_i-1)!}\frac{\partial^{m_i-1}}{\partial \nu^{m_i-1}}\psi_i
                   \end{pmatrix}^T\Big|_{\nu=\nu_j},
$$
$$\widehat{\phi_j}=\begin{pmatrix}
                     \phi_j, & \frac{\partial}{\partial \mu}\phi_j, & \cdots,  &\frac{1}{(m_j-1)!}\frac{\partial^{m_j-1}}{\partial \mu^{m_j-1}}\phi_j  \\
                   \end{pmatrix}\Big|_{\mu=\mu_i},
$$
$$P^{[ij]}=\left(P^{[ij]}_{kl}\right)_{m_{k},m_l},$$
$$P^{[ij]}_{kl}=\frac{1}{(k-1)!(l-1)!}\frac{\partial^{k+l-2}}{\partial
\nu^{k-1}\partial\mu^{l-1}}\left(\frac{\Psi_i(\nu)\sigma_3\Phi_j(\mu)}{\mu+\nu}-\frac{\Psi_i(\nu)\Phi_j(\mu)}{\mu-\nu}\right)\Big|_{\mu=\mu_i,\nu=\nu_j},$$
and
\begin{equation*}
   Q_1=\begin{pmatrix}
           Q^{[11]} & Q^{[12]} & \cdots & Q^{[1r]} & \widehat{\chi_1} \\
          Q^{[21]} & Q^{[22]} & \cdots & Q^{[2r]} &  \widehat{\chi_2} \\
          \vdots & \vdots&\ddots  & \vdots & \vdots \\
          Q^{[r1]} & Q^{[r2]} & \cdots & Q^{[rr]} & \widehat{\chi_r} \\
          \widehat{\varphi_1} & \widehat{\varphi_2} & \cdots  & \widehat{\varphi_r} & 0  \\
        \end{pmatrix},\;\;Q=\begin{pmatrix}
           Q^{[11]} & Q^{[12]} & \cdots & Q^{[1r]}  \\
          Q^{[21]} & Q^{[22]} & \cdots & Q^{[2r]}  \\
          \vdots & \vdots&\ddots  & \vdots \\
          Q^{[r1]} & Q^{[r2]} & \cdots & Q^{[rr]}  \\
        \end{pmatrix},
\end{equation*}
with
$$\widehat{\chi_i}=\begin{pmatrix}
                     \chi_i, &
                     \frac{\partial}{\partial \nu}\chi_i, &
                     \cdots, &
                     \frac{1}{(m_i-1)!}\frac{\partial^{m_i-1}}{\partial \nu^{m_i-1}}\chi_i
                   \end{pmatrix}^T\Big|_{\nu=\nu_i},
$$
$$\widehat{\varphi_j}=\begin{pmatrix}
                     \varphi_j, & \frac{\partial}{\partial \mu}\varphi_j, & \cdots,  &\frac{1}{(m_j-1)!}\frac{\partial^{m_j-1}}{\partial \mu^{m_j-1}}\varphi_j  \\
                   \end{pmatrix}\Big|_{\mu=\mu_j},
$$
$$Q^{[ij]}=\left(Q^{[ij]}_{kl}\right)_{m_{k},m_l},$$
$$Q^{[ij]}_{kl}=-\frac{1}{(k-1)!(l-1)!}\frac{\partial^{k+l-2}}{\partial
\nu^{k-1}\partial\mu^{l-1}}\left(\frac{\Psi_i(\nu)\sigma_3\Phi_j(\mu)}{\mu+\nu}+\frac{\Psi_i(\nu)\Phi_j(\mu)}{\mu-\nu}\right)\Big|_{\mu=\mu_i,\nu=\nu_j}.$$


According to above theorems, it is not difficult to see that the
reductions \eqref{reduction1} and \eqref{reduction2} are still valid
for gDT-I and gDT-II. For the gDT-II, to reduce system
\eqref{lin} to DNLS, we must set $\mu_i=\nu_i^*$.

\section{High order solutions for DNLS}
Integrable nonlinear partial differential equations are well known
for their richness of solutions. To construct those solutions,
a number of approaches have been proposed including Inverse Scattering Transform (IST), Dressing
 method, Hirota's bilinear theory and Darboux (B\"{a}cklund) method, etc.

While classical DT is known to be a convenient tool to construct
N-soliton solutions, it can not be directly used to obtain the high
order solutions, which correspond to multiple poles of the
reflection  coefficient in the IST terminology (see \cite{VY} and
the references there). We will show in this section that the gDT
derived above can be applied to obtain various solutions for DNLS.
Indeed, apart from the high order solutions, a kind of new N-soliton
solutions will also appear. In subsection 4.1, we consider the
solutions with the VBC. N-rational solitons, high order rational
solitons and high order solitons are worked out. In subsection 4.2,
we construct the high order solutions with NVBC which include the
high order rational solutions with NVBC and high order rogue wave
solutions.


\subsection{Solutions with VBC}
Applying DT-I to vacuum, we may obtain three kinds of solutions, namely plane wave solutions, N-phase
solutions (periodic  solutions) and N-soliton solutions (see \cite{XHW}). Additionally, if we take limit of
the soliton solutions, we can find rational solutions \cite{KN,XHW}.
In this subsection, we consider the N-rational solutions first.  As we
pointed out, the rational solitons are the limit cases to the soliton
solutions. The different behaviors of the
high order rational solitons and high order solitons are indicated.

In the first two cases, gDT-I will be used, while for the case
3, it is more convenient to use the gDT-II since the spectral parameters need be conjugated each other.

$\mathbf{Case\; 1}$: N-rational solutions.

We first consider the rational solitons and their higher order analogies. 
For the seed solution $u=0$, the special solution
for Lax pair \eqref{lin} with the reduction $u=-v^*$ is
\begin{equation}\label{phi-psi}
    \begin{pmatrix}
      \varphi \\
      \phi \\
    \end{pmatrix}
    =\begin{pmatrix}
           e^{-{\rm i}\zeta^{-2}(x+2\zeta^{-2}t+c)} \\
           e^{{\rm i}\zeta^{-2}(x+2\zeta^{-2}t+c)}
         \end{pmatrix},
\end{equation}
where $c$ is a complex constant. For higher order solutions,  $c$
will be taken as a polynomial function of $\zeta$ so that the high order solutions
  with free parameters may be constructed.

\begin{figure}
          \centering
       \includegraphics[width=5in]{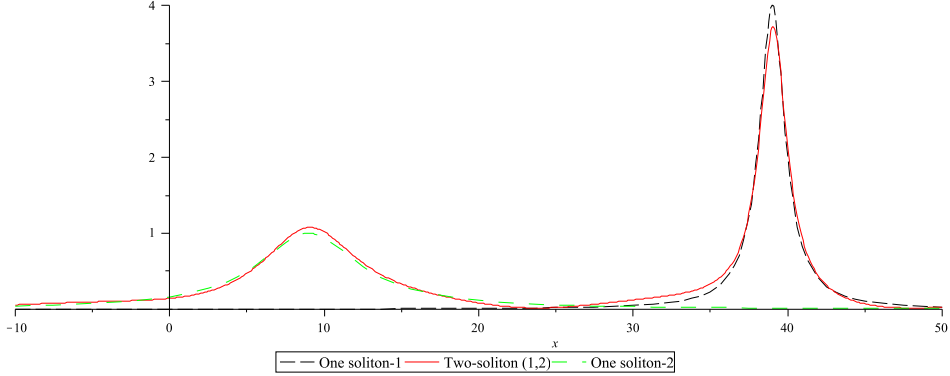}
\caption{Two-soliton (1,2) with the parameters $\zeta_1=2{\rm i}$,
$\zeta_2=4{\rm i}$ and $c_1=c_2=1$; One soliton-1 with parameters
$\zeta_1=4{\rm i}$ and $c_1=1$; One soliton-2 with parameters
$\zeta_1=2{\rm i}$ and $c_1=1$. 
}
\label{p1}
      \end{figure}

To obtain the N-rational solitons, we introduce vectors
\begin{equation*}
    y=[\varphi,\zeta\phi,\cdots,\zeta^{2N-2}\varphi,\zeta^{2N}\varphi],\quad
    z=[\varphi,\zeta\phi,\cdots,\zeta^{2N-2}\varphi,\zeta^{2N-1}\phi],
\end{equation*}
and define the matrices
\begin{equation*}
  Y=\begin{pmatrix}
        y_1 \\
        y_1^{(1)} \\
        \vdots \\
        y_N \\
        y_N^{(1)} \\
      \end{pmatrix}, \quad  Z=\begin{pmatrix}
        z_1 \\
        z_1^{(1)} \\
        \vdots \\
        z_N \\
        z_N^{(1)} \\
      \end{pmatrix},
\end{equation*}
where $y_{i}=y|_{\zeta=\zeta_i, c=c_i}$ and $z_{i}=z|_{\zeta=\zeta_i,c=c_i}$, the superscript $^{(1)}$ represents the first derivative
to $\zeta$. Then the N-rational soliton can be represented as
\begin{equation}\label{mrational}
    u[N]=-\left(\frac{\det{Y}}{\det{Z}}\right)_x.
\end{equation}
Taking $\zeta_1={\rm i}a$, we have
\begin{equation*}
    u[1]=\frac{4a^3[4{\rm i}(a^2x-4t+a^2c)-a^4]e^{\frac{2{\rm i}(a^2x-2t+a^2c)}{a^4}}}{[4{\rm
    i}(a^2x-4t+a^2c)+a^4]^2},
\end{equation*}
which appeared already in  \cite{XHW}.
The velocity for this rational soliton is $a^2/4$ and the center is
along the line $a^2x-4t+a^2c=0$. The altitude for $|u[1]|^2$ is
$16/a^2$. A simple analysis shows that this 2-rational soliton does not
possess phase shift when $t\rightarrow \pm\infty$, which is different
from the 2-soliton of NLS. This phenomenon is illustrated by Fig. \ref{p1}.

$\mathbf{Case\; 2}$: High order rational solitons

\begin{figure}
        \subfigure[1-st rational soliton]{
        \begin{minipage}[b]{0.4\textwidth}
          \centering
       \includegraphics[width=2.8in]{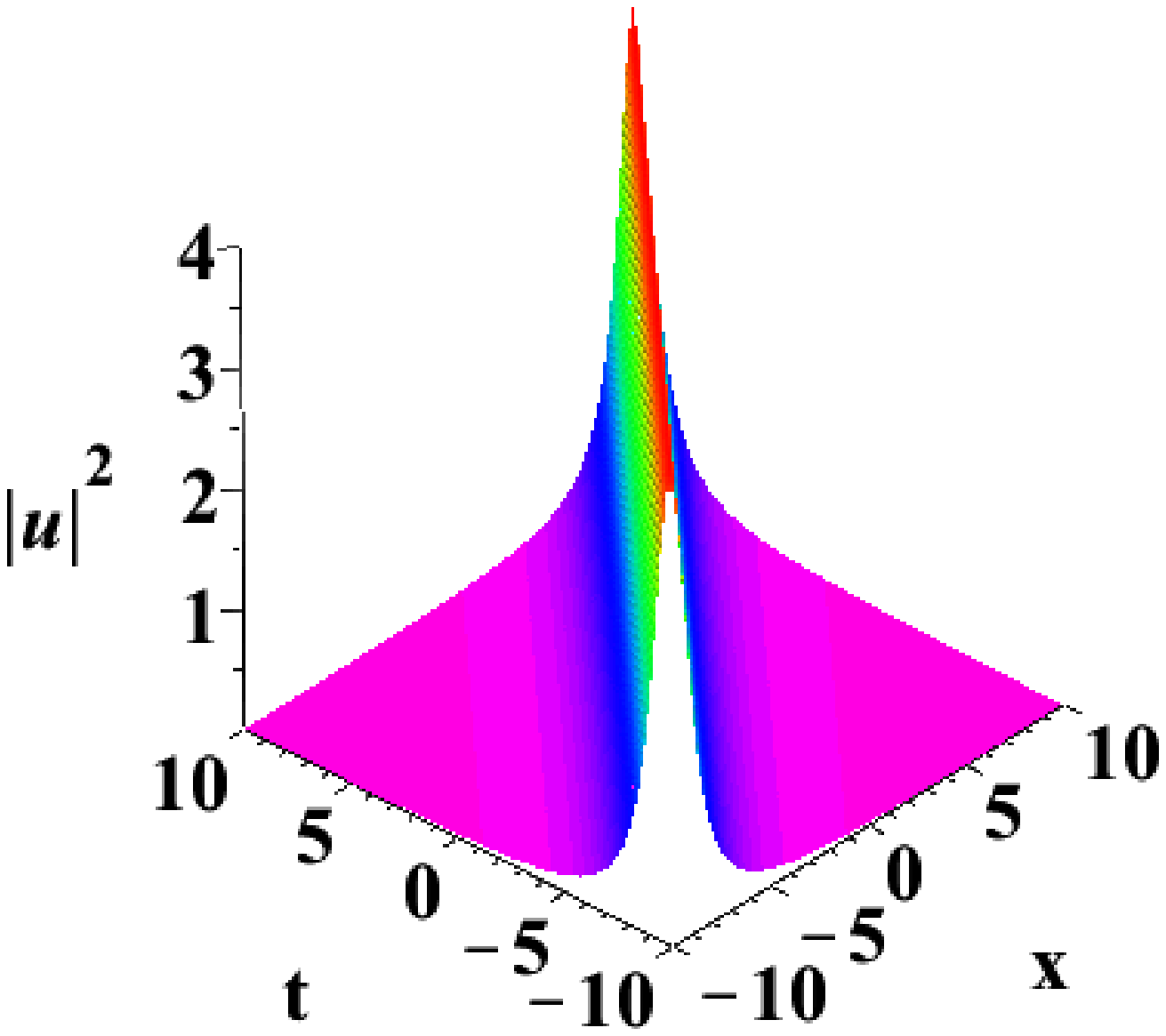}
        \end{minipage}}%
        \hspace{0.04\textwidth}%
       \subfigure[2-rd rational soliton]{ \begin{minipage}[b]{0.4\textwidth}
          \centering
       \includegraphics[width=3in]{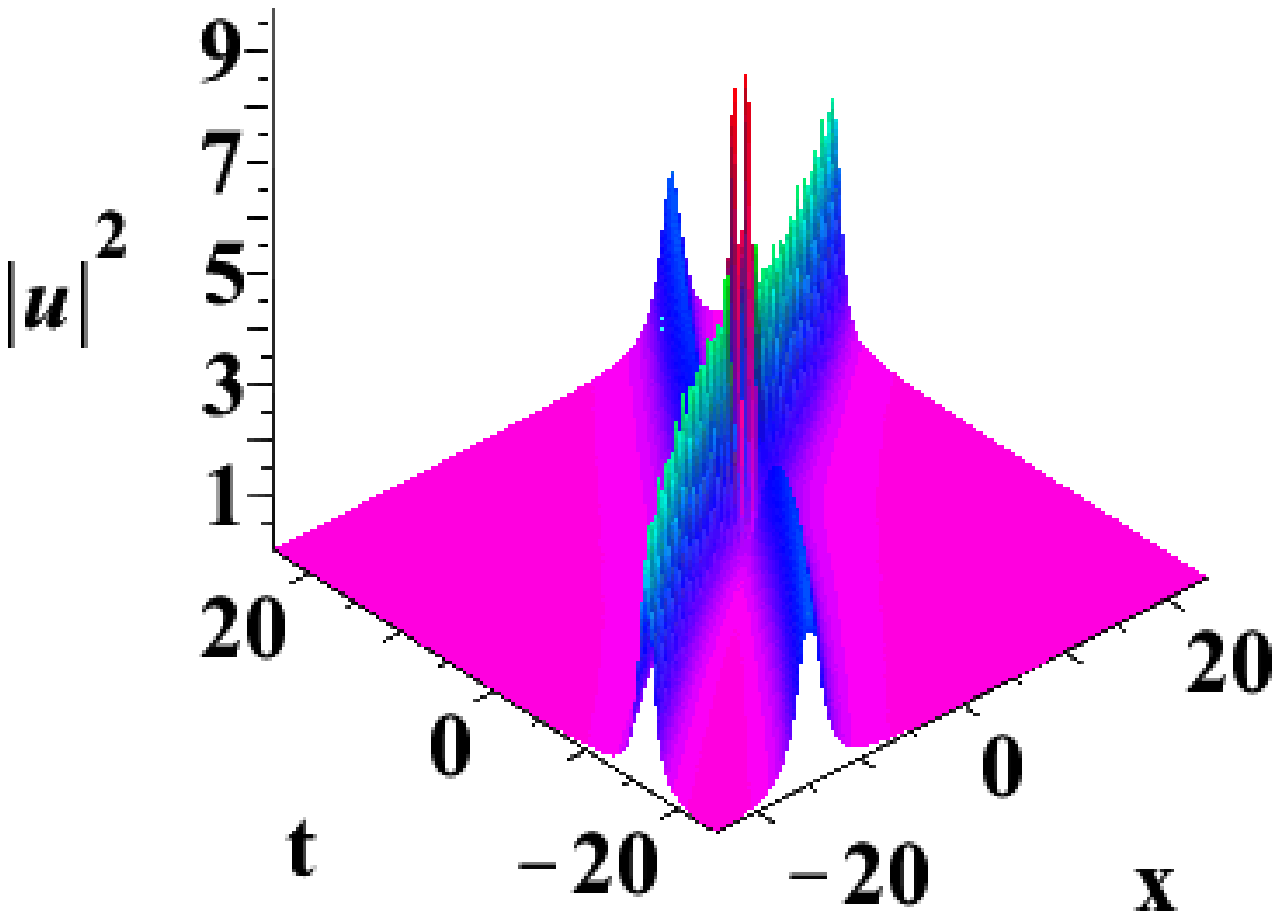}
        \end{minipage}}
\caption{(Color online): High order rational solution with VBC: The
parameters $\zeta_1=2{\rm i}$ and $c_1=0$}\label{p2}
      \end{figure}

Next we consider the high order rational solutions. Set the matrices
\begin{equation*}
  Y_1=\begin{pmatrix}
        y_1 \\
        y_1^{(1)} \\
        \vdots \\
        y_1^{(2N-1)} \\
      \end{pmatrix}, \quad  Z_1=\begin{pmatrix}
       z_1 \\
        z_1^{(1)} \\
        \vdots \\
        z_1^{(2N-1)} \\
      \end{pmatrix},
\end{equation*}
where the superscript $^{(i)}$ represents the $i$-th derivative with
respect to $\zeta$. It follows that the $N-$order rational soliton
for DNLS with VBC can be formulated as
\begin{equation}\label{mrationa2}
    u[N]=-\left(\frac{\det Y_1}{\det Z_1}\right)_x.
\end{equation}
By choosing appropriate parameters, we have the first and second order rational
solitons with VBC, which are plotted in Fig. \ref{p2}.

$\mathbf{Case \; 3}$: High order solitons

To obtain the high order soliton solutions, we start with the seed
solution $u[0]=0$. The special solution for Lax pair \eqref{lin}
with the reduction $u=-v^*$ at $u=u[0]$ is
\begin{equation*}
    \begin{pmatrix}
      \varphi \\
      \phi \\
    \end{pmatrix}
    =\begin{pmatrix}
           e^{-{\rm i}\mu^{-2}(x+2\mu^{-2}t+c)} \\
           e^{{\rm i}\mu^{-2}(x+2\mu^{-2}t+d)} \\
         \end{pmatrix},
\end{equation*}
and the special solution for conjugate Lax pair \eqref{conlin} with
the reduction $u=-v^*$ at $u=u[0]$ reads as
\begin{equation*}
    \begin{pmatrix}
   \chi,&\psi
    \end{pmatrix}=\begin{pmatrix}
           e^{{\rm i}\nu^{-2}(x+2\nu^{-2}t+c^*)},&
           e^{-{\rm i}\nu^{-2}(x+2\nu^{-2}t+d^*)}
         \end{pmatrix}.
\end{equation*}
Then, the N-th order soliton solution for DNLS is given by
\begin{equation}\label{N-order}
    u[N]=-\left(\frac{\det M_1}{\det M}\right)_x,\quad M=(M_{ij})_{N\times N}
\end{equation}
where \begin{eqnarray*}
  M_{ij}=\frac{\mathrm{d}^{i+j-2}}{\mathrm{d}\nu^{i-1}\mathrm{d}\mu^{j-1}}\frac{2[\nu e^{{\rm i}(\nu^{-2}-\mu^{-2})[x+2(\nu^{-2}+\mu^{-2})t]}+\mu
  e^{-{\rm
  i}(\nu^{-2}-\mu^{-2})[x+2(\nu^{-2}+\mu^{-2})t]}]}{\nu^2-\mu^2}|_{\nu=\mu^*},
      \end{eqnarray*}
\begin{equation*}
    M_1=\begin{pmatrix}
          M & Y^T \\[10pt]
          X & 0 \\
        \end{pmatrix},\quad  X=[
\varphi,\frac{\mathrm{d}}{\mathrm{d}\mu}\varphi,\cdots,\frac{\mathrm{d}^{N-1}}{\mathrm{d}\mu^{N-1}}\varphi],\quad
Y=[\psi,\frac{\mathrm{d}}{\mathrm{d}\nu}\psi,\cdots,\frac{\mathrm{d}^{N-1}}{\mathrm{d}\nu^{N-1}}\psi].
       \end{equation*}

\begin{figure}
        \subfigure[2-nd order soliton]{
        \begin{minipage}[b]{0.4\textwidth}
          \centering
       \includegraphics[width=3in]{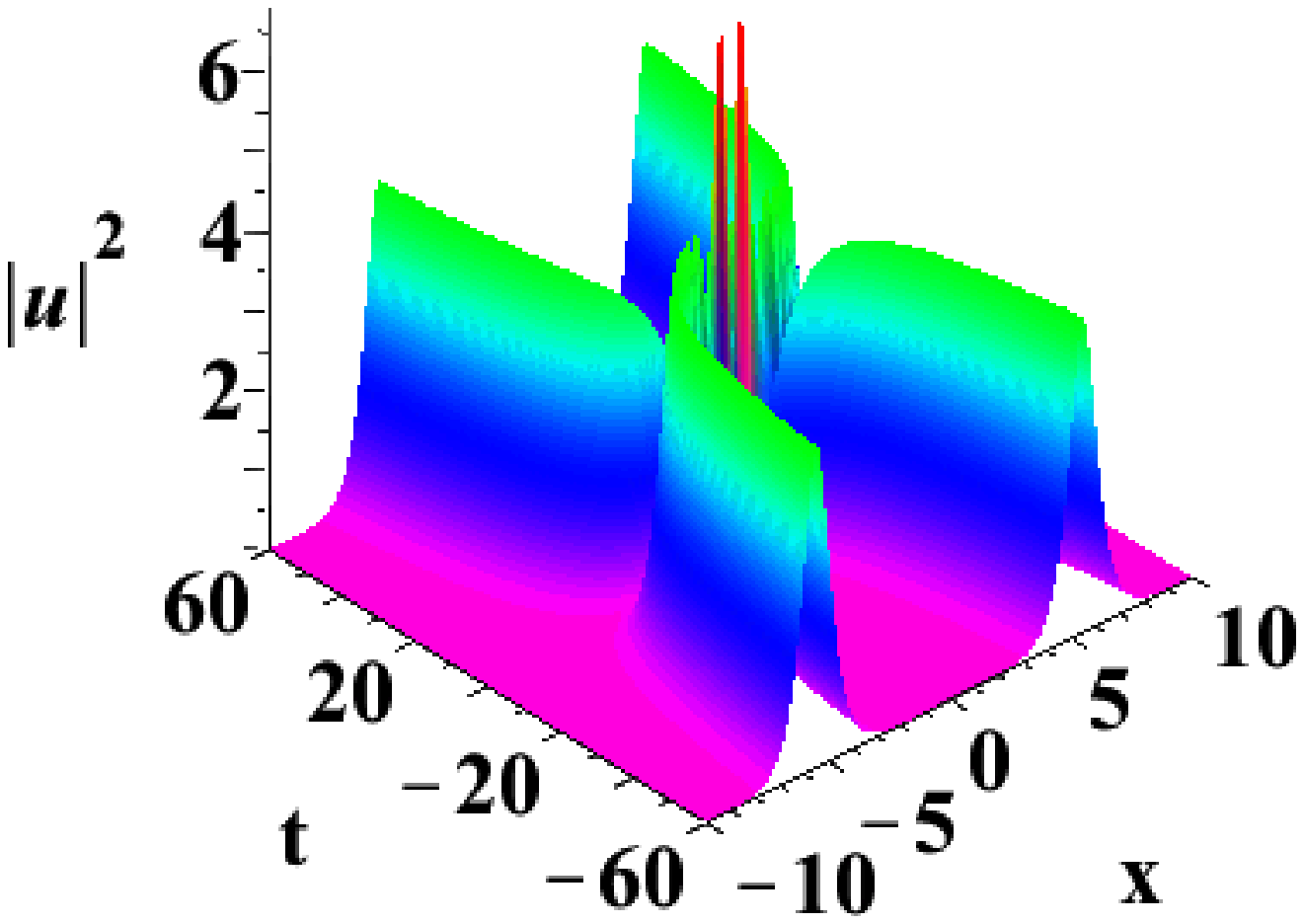}
        \end{minipage}}%
        \hspace{0.04\textwidth}%
       \subfigure[3-rd order soliton]{ \begin{minipage}[b]{0.4\textwidth}
          \centering
       \includegraphics[width=3in]{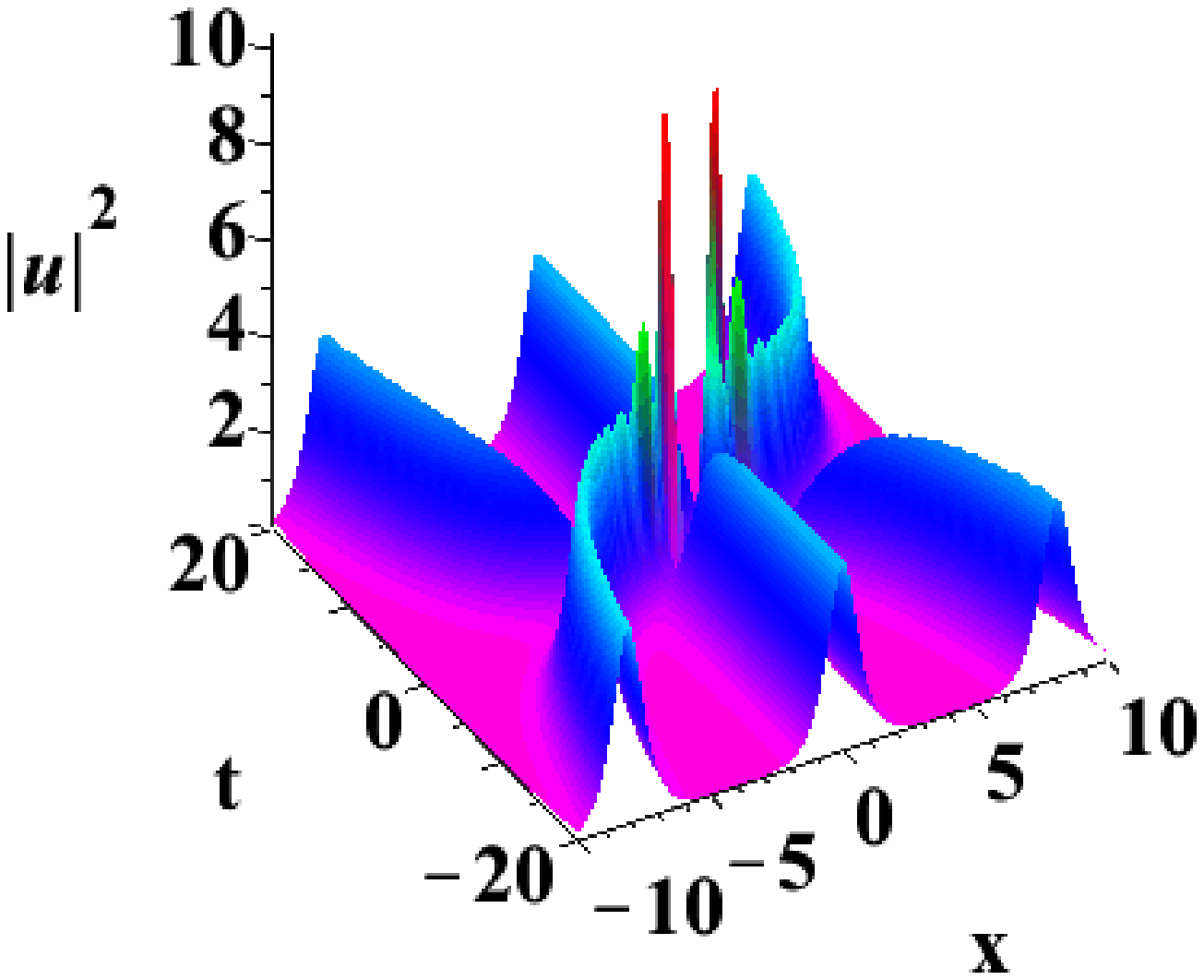}
        \end{minipage}}
\caption{(Color online): High order soliton with VBC: The parameters
$\mu=1+{\rm i}$ and $c=d=0$}\label{p5}
      \end{figure}

The second order and third order soliton solutions are shown in Fig.
\ref{p5}. The high order soliton with more free parameters may be
obained if $c$ or $d$ are taken as polynomial functions of $\mu$.

\subsection{Solution with NVBC}
The solutions with NVBC may be obtained by applying DT to zero solution.
As illustrated in \cite{S}, one fold DT-I could be
used to yield the plane wave solution. Thus  we will consider
the high order rational soliton solutions resulted from vacuum first.
To find more general solutions with NVBC, we may apply DT to the general plane wave solution. This will be considered in case
2 and the genuine rational solutions and their high order analogies with NVBC will be calculated. In case 3, we construct high order
rogue wave solutions. As above, gDT-I will be employed in first two cases and gDT-II will be adopted in the last case.

To the high order rational solution with NVBC and high order rogue
waves, because the all locate in the branch point spectral. We must
use some tricks to deal with this problem. Besides the high order
rogue wave solution, the high order breather solution and periodic
solution can be readily to obtain similarly. Because they are
nothing but using the the formula \eqref{filedsn1} directly like the
above subsection: case 3. Thus we omit them in our work.

$\mathbf{Case \;1}:$ High order rational solitons with NVBC from
vacuum

\begin{figure}
        \subfigure[1-st order rational soliton]{
        \begin{minipage}[b]{0.4\textwidth}
          \centering
       \includegraphics[width=3in]{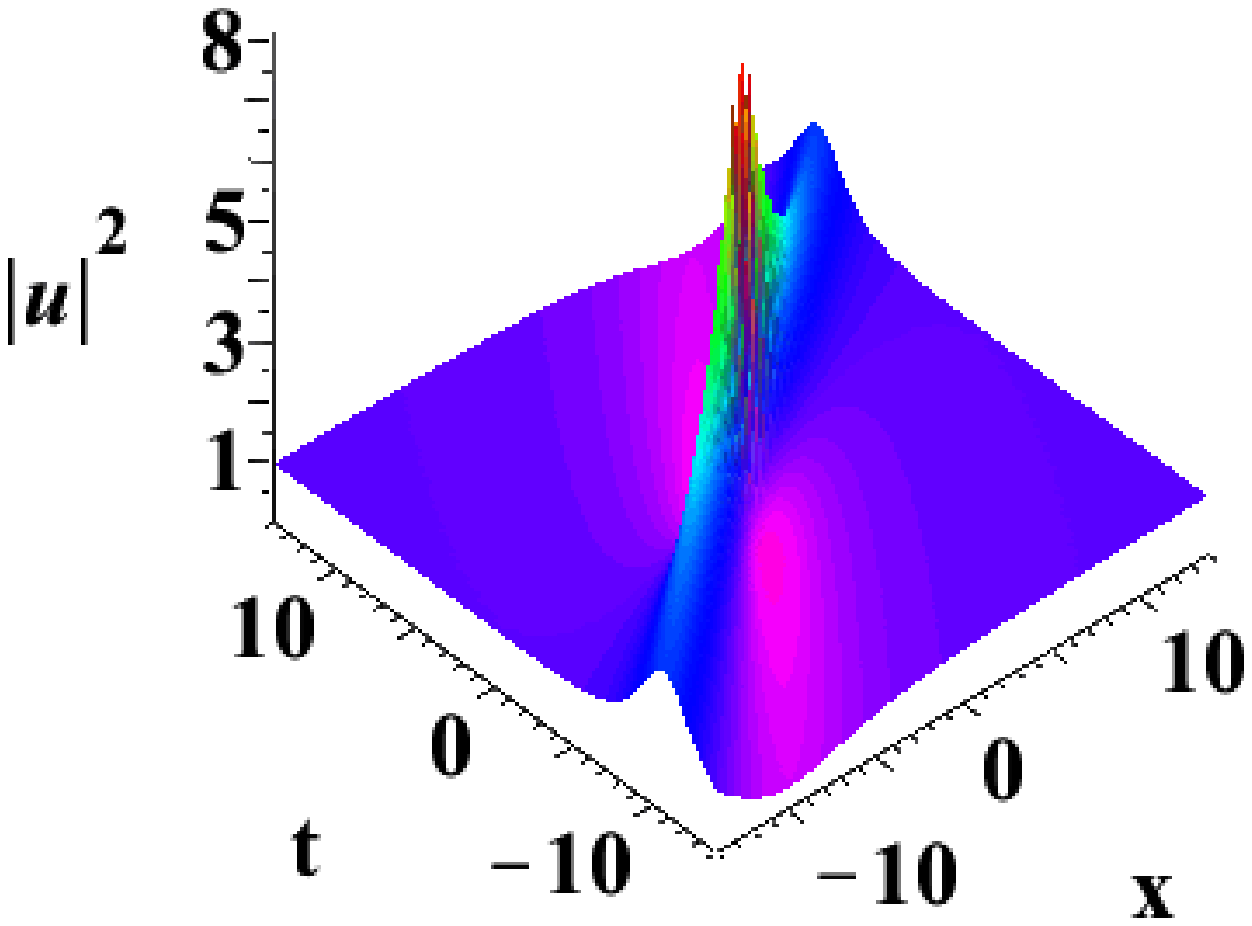}
        \end{minipage}}%
        \hspace{0.04\textwidth}%
       \subfigure[2-nd order rational soliton]{ \begin{minipage}[b]{0.4\textwidth}
          \centering
       \includegraphics[width=3in]{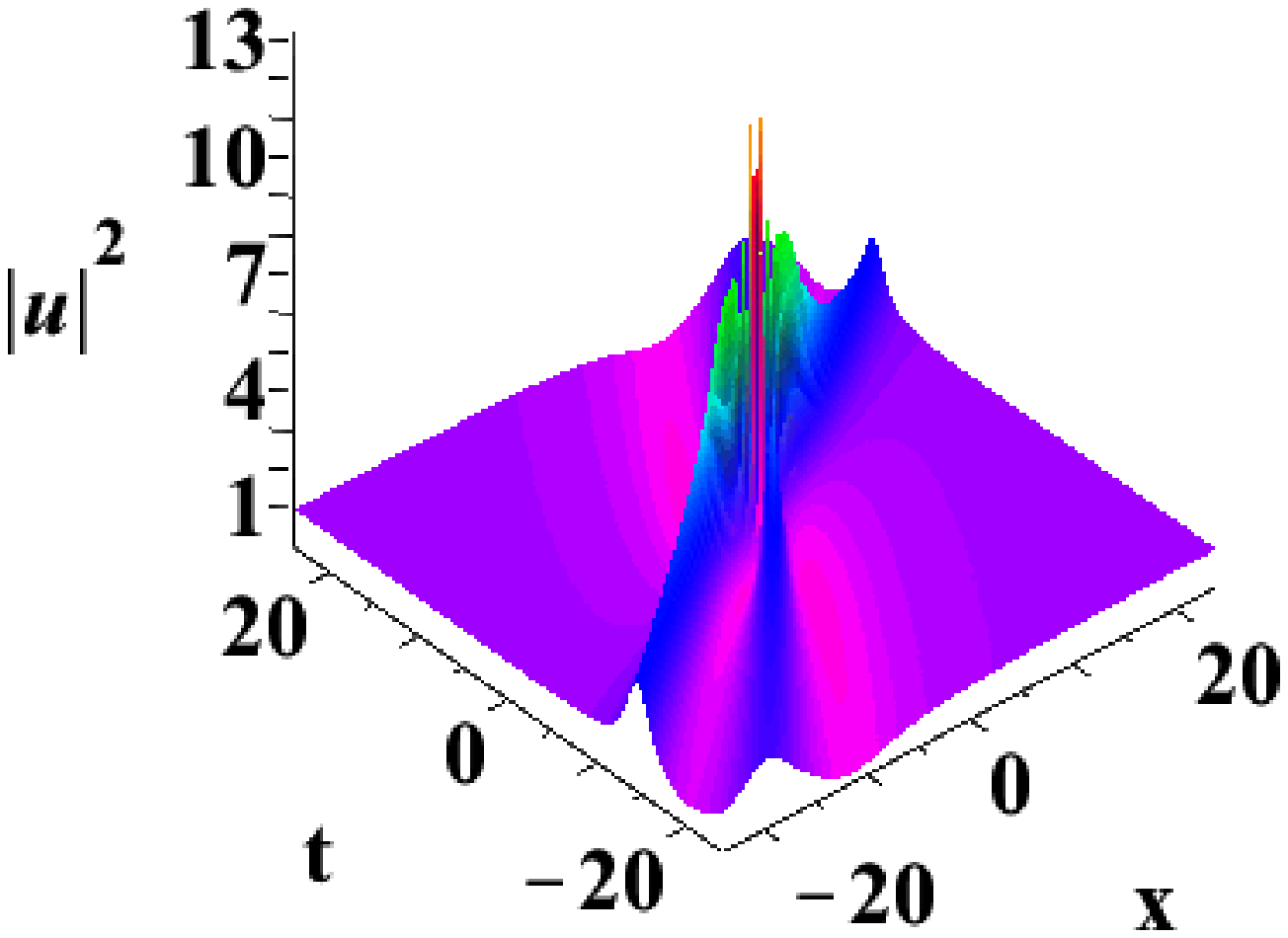}
        \end{minipage}}
\caption{(Color online): Rational solitons with NVBC:
The parameters $\zeta_1=2{\rm i}$ and $c_1=0$}\label{p3}
      \end{figure}

To obtain high order rational solutions with NVBC, the order of
determinants must be odd. Define the matrices
\begin{equation*}
  \widehat{Y}_1=\begin{pmatrix}
        \widehat{y_1} \\
        \widehat{y_1}^{(1)} \\
        \vdots \\
        \widehat{y_1}^{(2N)} \\
      \end{pmatrix}, \quad  \widehat{Z}_1 =\begin{pmatrix}
       \widehat{z_1} \\
        \widehat{z_1}^{(1)} \\
        \vdots \\
        \widehat{z_1}^{(2N)} \\
      \end{pmatrix},
\end{equation*}
where
\begin{equation*}
    \widehat{y_1}=[\varphi,\zeta_1\phi,\cdots,\zeta_1^{2N-1}\phi,\zeta_1^{2N+1}\phi],
\quad
  \widehat{z_1}=[\varphi,\zeta_1\phi,\cdots,\zeta_1^{2N-1}\phi,\zeta_1^{2N}\varphi],
\end{equation*}
and $\phi$, $\psi$ are given by \eqref{phi-psi}. Then the high order
rational solitons with NVBC can be represented as
\begin{equation}\label{mrationa2}
    u[N]=-\left(\frac{\det\widehat{ Y}_1}{\det\widehat{Z}_1}\right)_x.
\end{equation}
The first and second order rational solitions in NVBC are shown in
Fig. \ref{p3}.

$\mathbf{Case\; 2}:$ High order rational solitons with NVBC from
plane wave solution

To construct these solutions, we take the seed solution as
\[
u[0]=A\exp(2{\rm
i}\theta_2),
\]
where $ \theta_2=\frac{1}{2}[ax-(A^2a+a^2)t+c],  c\in
\mathbb{R}$.
The corresponding fundamental solution for Lax pair \eqref{lin} is
\begin{equation}\label{fund}
    \Phi=\begin{pmatrix}
           e^{{\rm i}(\theta_1+\theta_2)+\phi} & e^{-{\rm i}(\theta_1-\theta_2)-\phi} \\
           e^{{-\rm i}(\theta_1+\theta_2)+\phi} & e^{{\rm i}(\theta_1-\theta_2)-\phi} \\
         \end{pmatrix},
\end{equation}
where
$$\theta_1=\frac{1}{2}\arccos{(-\frac{2+a\zeta^2}{2{\rm i}A\zeta})},$$
and
$$\phi=\frac{1}{2}\sqrt{-(2\zeta^{-2}+a)^2-4A^2\zeta^{-2}}[x-(a+A^2-2\zeta^{-2})t+d].$$
To resolve the reduction \eqref{reduction1}, we assume
$-4A^2\zeta^2-(2+a\zeta^2)^2>0$, $d\in \mathbb{R}$, $\zeta\in {\rm
i}\mathbb{R}$. The corresponding dark solitons and bright solitons were studied and analyzed in \cite{S,XHW} in details.
The authors of \cite{S} also illustrated  certain limit cases, but they did not give the explicit expression for those solutions.


 In the case $-4A^2\zeta^2-(2+a\zeta^2)^2=0$, $\Phi$ given by the formula \eqref{fund}, which does not qualify as
the fundamental solution,  is in fact a constant. Thus the gDT could not generate interesting solutions.
 To obtain meaningful solutions, we must find another solution for \eqref{lin}.
To this end, we turn to the limit technique.

For convenience, we consider the special case $a=c=0, A=1$ which leads to the genuine rational solutions.
We will expand the solution for Lax pair at $\zeta={\rm i}$.
With the special solution
$$\Phi_1=\frac{\Phi|_{\zeta={\rm i}(1+f)}C
}{f^{1/2}},\quad C=\begin{pmatrix}
                                                 1 \\
                                                 1 \\
                                               \end{pmatrix},$$ at $f=0$, we have
\begin{eqnarray*}
\Phi_1&=&Y_0+Y_1f+\cdots+Y_nf^n+\cdots,\\
 Y_n&=&\begin{pmatrix}
                                                           x_n\\
                                                           y_n\\
                                                         \end{pmatrix}=\lim_{f=0}\frac{1}{n!}\frac{\mathrm{d}^n}{\mathrm{d}f^{n}}\Phi_1(f).
\end{eqnarray*}
The high order genuine rational solution in NVBC is represented as
the following:
\begin{enumerate}
  \item
When $N=2l-1$, we have
\begin{equation}\label{nvbc-o}
u[2l-1]=-1-\left[\frac{\det(B)}{\det(A)}\right]_x,
\end{equation}
 where
$$A_{i,2j-1}=B_{i,2j-1}={\rm i}^{2j-1}\sum_{k=0}^{\min(i-1,2j-2)}C_{2j-1}^kx_k,\quad(j=1,2,\cdots,l-1),\quad B_{i,2l-1}={\rm i}^{2l}\sum_{k=0}^{i-1}C_{2l}^kx_k,$$
$$A_{i,2j}=B_{i,2j}={\rm i}^{2j}\sum_{k=0}^{\min(i-1,2j-1)}C_{2j}^ky_k,\quad C_{m}^n=\frac{m!}{n!(m-n)!}.$$
 \item
When $N=2l$, we have
\begin{equation}\label{nvbc-e}
u[2l]=1-\left[\frac{\det(D)}{\det(C)}\right]_x,
\end{equation}
 where
$$C_{i,2j-1}=D_{i,2j-1}={\rm i}^{2j-1}\sum_{k=0}^{\min(i-1,2j-2)}C_{2j-1}^kx_k,\quad C_{i,2j}={\rm i}^{2j}\sum_{k=0}^{\min(i-1,2j-1)}C_{2j}^ky_k, \quad(j=1,2,\cdots,l)$$
$$D_{i,2k}=C_{i,2k},\quad( k=1,2,\cdots,l-1),\quad D_{i,2l}={\rm i}^{2l+1}\sum_{k=0}^{i-1}C_{2l+1}^ky_k.$$
\end{enumerate}
In particular, taking $d=ef$, where $e$ is real number, we have
\begin{eqnarray*}
  x_0 &=& \sqrt{2}(2x-6t-{\rm i}),\quad  y_0 = \sqrt{2}(2x-6t+{\rm i}) , \\
  x_1 &=& \sqrt{2}\left[
                                    \frac{2}{3}x^3-6x^2t+18xt^2-18t^3-4x+20t+2e+{\rm i}(\frac{1}{2}-x^2+6xt-9t^2)\right], \\
  y_1 &=&\sqrt{2}\left[\frac{2}{3}x^3-6x^2t+18xt^2-18t^3-4x+20t+2e+{\rm
  i}(x^2-\frac{1}{2}-6xt+9t^2)\right].
\end{eqnarray*}
Thus by means of above formula \eqref{nvbc-o}, the first order genuine rational soliton
solution reads
\begin{equation*}
    u[1]=-\frac{(-2x+6t-{\rm i})(-2x+6t+3{\rm i})}{(-2x+6t+{\rm
    i})^2}.
\end{equation*}
Similarly,  \eqref{nvbc-e} provides us the following second-order genuine rational soliton with NVBC
\begin{equation*}
     u[2]=\frac{L_1^*L_2}{L_1^2},
\end{equation*}
where the superscript $^*$ stands for complex conjugation and
\begin{eqnarray*}
L_1 &=& -8x^3+72x^2t-216xt^2+216t^3+12{\rm i}x^2-72{\rm i}xt+108{\rm
i}t^2-18x+102t+3 {\rm i}+12
  e,\\
   L_2 &=& -8x^3+72x^2t-216xt^2+216t^3+36{\rm i}x^2-216{\rm i}xt+324{\rm i}t^2+30x-42t+12 e-15 {\rm
   i}.
\end{eqnarray*}

%
%

$\mathbf{Case\; 3}$ : High-order rogue wave solutions

The rogue wave solution for DNLS was first derived in \cite{XHW} via
DT, to the best of our knowledge. However, the classical DT can be not be used directly to obtain high order rogue wave solutions.
According to above, we can see that the gDT is a very efficient way to
obtain high order solutions.

\begin{figure}
        \subfigure[]{
        \begin{minipage}[b]{0.4\textwidth}
          \centering
       \includegraphics[width=3in]{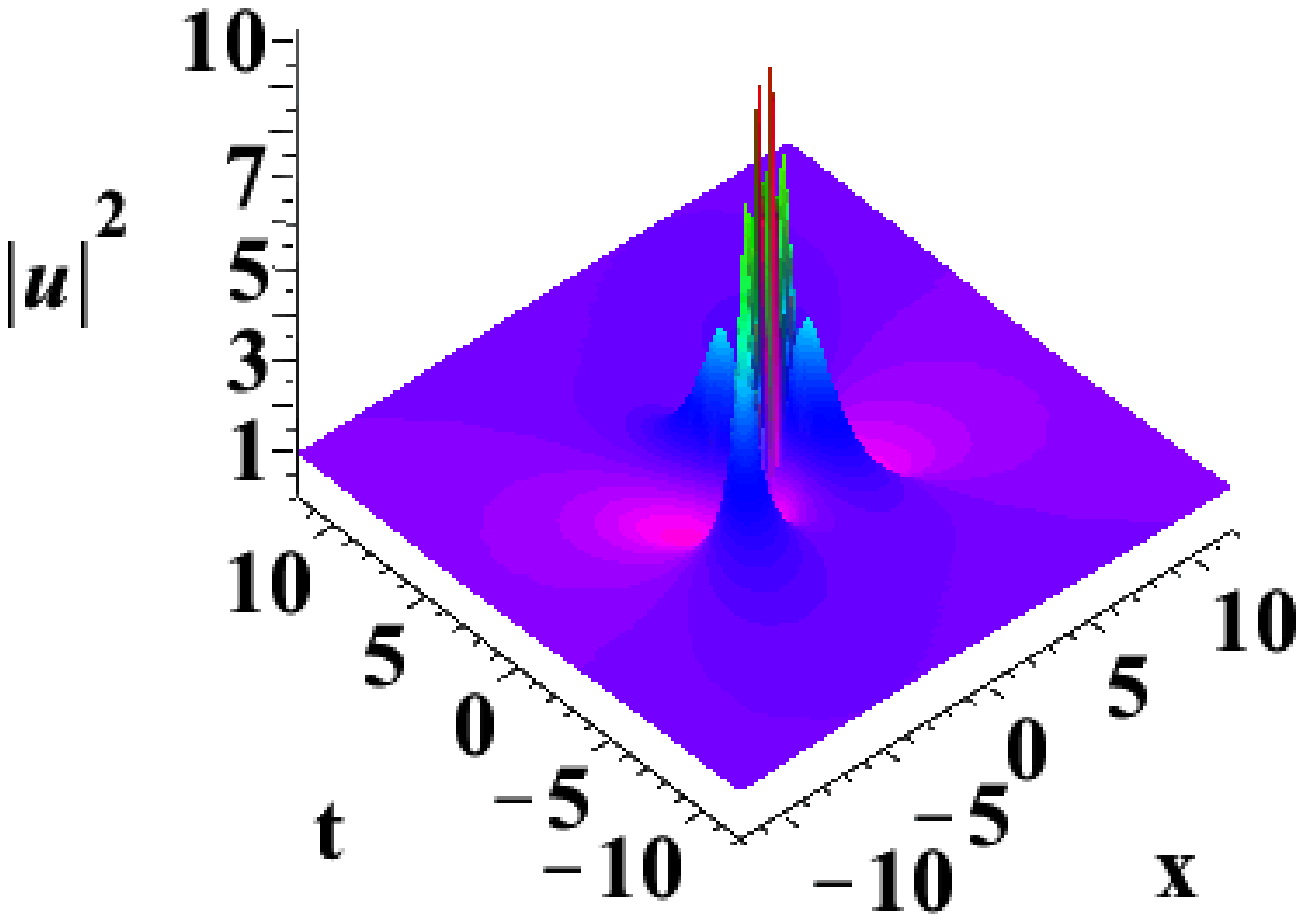}
        \end{minipage}}%
        \hspace{0.04\textwidth}%
       \subfigure[]{ \begin{minipage}[b]{0.4\textwidth}
          \centering
       \includegraphics[width=3in]{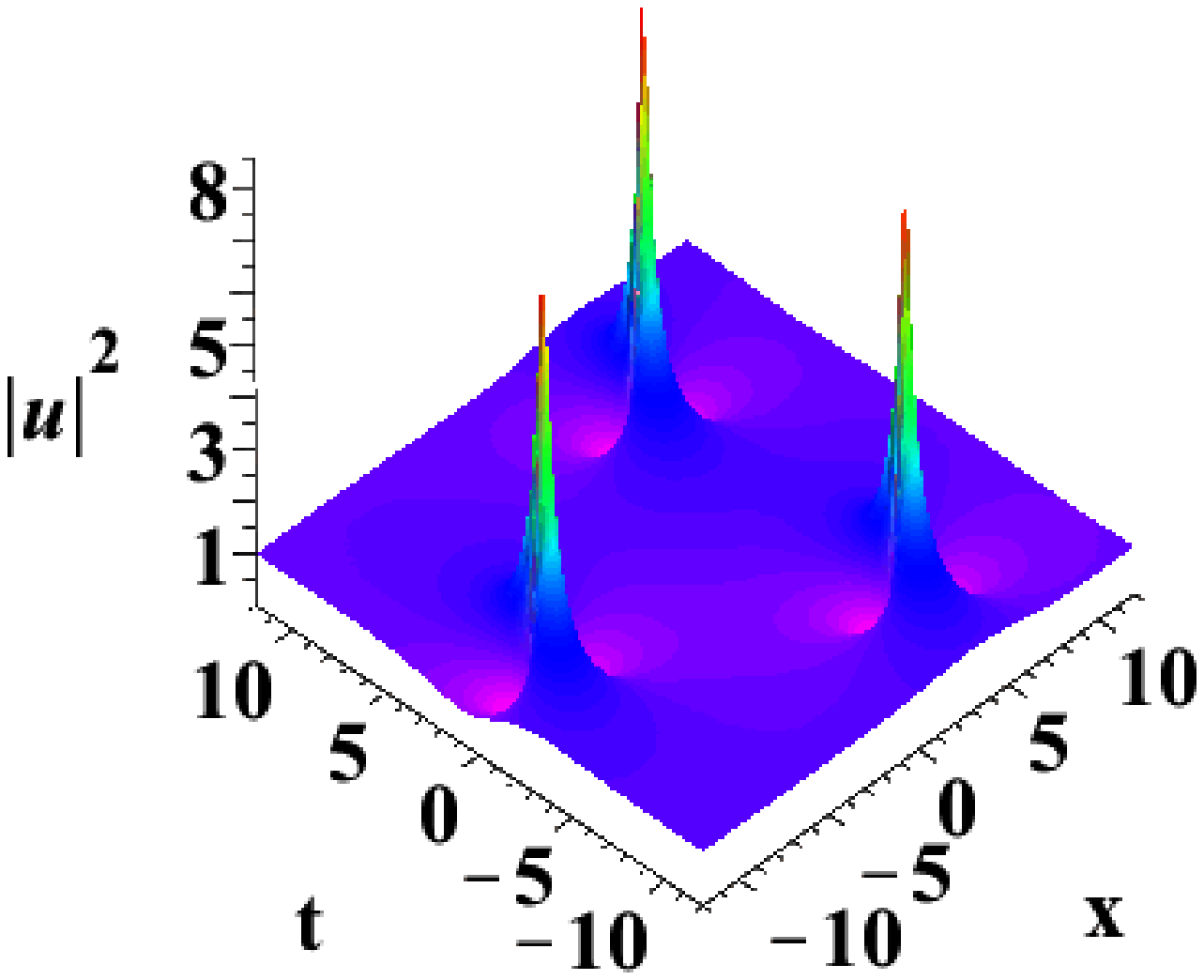}
        \end{minipage}}
\caption{(Color online):   Second order rogue wave with the parameters (a)
$e=g=0$; (b)   $e=0$ and $g=100$.}\label{p7}
      \end{figure}

In order to get the higher-order rogue wave solutions, for
simplicity,  we consider the seed solution $u[0]=\exp(-{\rm i}x)$.
The corresponding fundamental-matrix solution for Lax pair  is
\begin{equation*}
    \Phi=E\begin{pmatrix}
                       \alpha & \alpha^{-1}\\
                        -\alpha^{-1} & -\alpha \\
                      \end{pmatrix}\begin{pmatrix}
                                     \beta & 0 \\
                                     0 & \beta^{-1} \\
                                   \end{pmatrix},\quad E=\begin{pmatrix}
           \exp{(-\frac{1}{2}{\rm i}x)} & 0 \\
           0 & \exp{(\frac{1}{2}{\rm i}x)} \\
         \end{pmatrix}
\end{equation*}
where
\[
\alpha=[(2\zeta)^{-1}(\lambda-2{\rm i}+{\rm
i}\zeta^2)]^{1/2},\;
\beta=\exp{\left[\frac{1}{2}\lambda\zeta^{-2}(x+2\zeta^{-2}t+F(\zeta))\right]},\; \lambda=(-4-\zeta^4)^{1/2}
\]
and $F(\zeta)$ is a polynomial function for $\zeta$. Like above
case, by means of the limit technique, we expand the special
solution at $\zeta=1+{\rm i}$
$$\Phi_1=\frac{\Phi|_{\zeta=(1+{\rm i})(1+f)}C
}{f^{1/2}},\quad C=\begin{pmatrix}
                                                 1 \\
                                                 1 \\
                                               \end{pmatrix},$$ at $f=0$, and find
\begin{eqnarray*}
\Phi_1&=&Y_0+Y_1f+\cdots+Y_nf^n+\cdots ,\end{eqnarray*} where
\begin{eqnarray*}
    Y_n&=&\begin{pmatrix}
           x_n\\
           y_n\\
        \end{pmatrix}=\lim_{f=0}\frac{1}{n!}\frac{\mathrm{d}}{\mathrm{d}f^{n}}\Phi_1(f).
\end{eqnarray*}
Expilcitly, we have
\begin{eqnarray*}
x_0&=&\exp{[-\frac{1}{2}{\rm i}x]}(2x-2{\rm i}t -1-{\rm i}), \\
y_0&=&\exp{[\frac{1}{2}{\rm i}x]}(2x-2{\rm i}t+ 1+{\rm i}),\\
  x_1 &=&\exp{[-\frac{1}{2}{\rm i}x]}\left[-\frac{1}{3}x^3+{\rm i}x^2t+xt^2-\frac{1}{3}{\rm i}t^3+\frac{1+{\rm i}}{2}x^2+(1-{\rm i})xt-\frac{1+{\rm i}}{2}t^2\right.\\
  &&\left.-\frac{1}{2}{\rm i}x-\frac{5}{2}x+\frac{13}{2}{\rm i}t-\frac{1}{2}t+2 e+\frac{1}{2}+2 {\rm i} g\right],  \\
  y_1 &=&\exp{[\frac{1}{2}{\rm i}x]}\left[-\frac{1}{3}x^3+{\rm i}x^2t+xt^2-\frac{1}{3}{\rm i}t^3-\frac{1+{\rm i}}{2}x^2-(1-{\rm i})xt+\frac{1+{\rm i}}{2}t^2\right.\\
  &&\left.-\frac{1}{2}{\rm i}x-\frac{5}{2}x+\frac{13}{2}{\rm i}t-\frac{1}{2}t+2 e-\frac{1}{2}+2 {\rm
  i}
  g\right].
\end{eqnarray*}
Therefore, the $N$-th order rogue wave can be represented as
following
\begin{equation}\label{N-rogue}
u[N]=\exp[-{\rm i}x]-\left(\frac{\det M_1 }{\det M}\right)_x,
\end{equation}
where $M=\left(M_{ij}\right)_{N\times N}$,
\begin{equation*}
    M_1=\begin{pmatrix}
          M_{11} & M_{12} & \cdots&M_{1N}& y_0 \\
          M_{21} & M_{22} & \cdots&M_{2N}& y_1 \\
          \vdots & \vdots&\ddots&\vdots& \vdots \\
             M_{N1} & M_{N2} &\cdots&M_{NN}& y_{N-1} \\
          x_0 & x_1 & \cdots  &x_{N-1} & 0  \\
        \end{pmatrix},
\end{equation*}
\begin{eqnarray*}
M_{ij}&=&\sum_{k=0,l=0}^{i-1,j-1}-(-\frac{1}{2})^{i+j-(k+l+1)}C_{i+j-(k+l+2)}^{i-k-1}\\
&&\cdot\left[Y_k^{\dag}\sigma_3 Y_l \left(1-{\rm
i}\right)^{i-k-1}\left(1+{\rm i}\right)^{j-l-1} +{\rm i}Y_k^{\dag}
Y_l \left(1+{\rm i}\right)^{i-k-1}\left(1-{\rm
i}\right)^{j-l-1}\right].
\end{eqnarray*}
In particular,  the first-order and the second order rogue wave solutions are given by
\begin{equation*}
    u[1]=-\frac{[2t^2+2x^2-3-2{\rm i}(x+3t)][2x^2+2t^2-2{\rm i}(x-t)]}
    {[2x^2+2t^2+1+2{\rm i}(x-t)]^2}\exp[-{\rm i}x],
\end{equation*}
and
\begin{equation*}
    u[2]=\frac{L_1^*L_2}{L_1^2}\exp[-{\rm i}x],
\end{equation*}
where
\begin{eqnarray*}
  L_1 &=& 72[e^2+g^2]+[48x^3-144xt^2+72{\rm i}x^2+144{\rm i}xt-72{\rm i}t^2-144x-72t+36{\rm i}]e\\
  &&+[-144x^2t+48t^3+72{\rm i}x^2-144{\rm i}xt-72{\rm i}t^2-72x+432t-36{\rm i}]g
  \\
  &&+8x^6+24x^4t^2+24x^2t^4+8t^6+24{\rm i}x^5-24{\rm i}x^4t+48{\rm i}x^3t^2-48{\rm i}x^2t^3\\
  &&+24{\rm i}xt^4-24{\rm i}t^5-12x^4+48x^3t-216x^2t^2+48xt^3+180t^4+48{\rm i}x^3\\
  &&-288{\rm i}xt^2-336{\rm i}t^3+90x^2-72xt+666t^2+54{\rm i}x-198{\rm i}t+9,\\
  L_2
  &=&72[e^2+g^2]+[48x^3-144xt^2-72{\rm i}x^2+432{\rm i}xt+72{\rm i}t^2+144x+216t-180{\rm i}]e\\
  &&+[-144x^2t+48t^3+216{\rm i}x^2+144{\rm i}xt-216{\rm i}t^2+216x+144t+36{\rm i}]g\\
  &&+8x^6+24x^4t^2+24x^2t^4+8t^6-24{\rm i}x^5-72{\rm i}x^4t-48{\rm i}x^3t^2-144{\rm i}x^2t^3-24{\rm i}xt^4\\
  &&-72{\rm i}t^5-60x^4-144x^3t-504x^2t^2-144xt^3-60t^4+48 {\rm i} x^3+288 {\rm i} x^2 t\\
  &&+576 {\rm i} x t^2-528 {\rm i} t^3-198 x^2+504 x t-486 t^2+90 {\rm i} x+414 {\rm i}
  t+45.
\end{eqnarray*}
These solutions are plotted in Fig. \ref{p7} with different
parameters.


\section{Conclusion and Discussion}

The theory of DT is developed and two generalized Darboux transformations, gDT-I and gDT-II, are constructed  for DNLS. With the help of them, two generalized determinant solution formulae are obtained for this physically relevant equation.
Moreover, high order solitons, high order rogue waves and rational
solutions are given explicitly. We remark that the gDT-II is still valid for N-component DNLS system. In addition,
the above formula can be easily modified and applied to so-called Fokas-Lenells equation \cite{L}.

As shown in Figure 5, the second order rogue waves exhibit dynamics
which varies according to the different values of the parameters. It
is interesting to study the dynamics of the general high order
solutions. Also, it is interesting to calculate the energy of the
high order solutions, which could be done directly. However, the calculations involved in is
very tedious. These questions may be answered efficiently by a detailed analysis of
DNLS in the framework of inverse scattering method.


\bigskip
\noindent
{\bf Acknowledgments}

This work is supported by the National
Natural Science Foundation of China (grant number:10971222) and the Fundamental Research Funds for Central
Universities.

\end{document}